\documentclass[letterpaper,prd,aps,nofootinbib]{revtex4}
\usepackage{amssymb}
\usepackage{bm}
\usepackage{enumerate}
\usepackage{amsmath}
\usepackage{feynmf}
\usepackage{slashed}
\usepackage{wrapfig}
\usepackage[caption=false]{subfig}
\usepackage{graphicx}
\usepackage{filecontents}
\usepackage{mathtools}
\usepackage{xcolor}

\newcommand\be{\begin{equation}}
\newcommand\ee{\end{equation}}

\begin{document}
\setlength{\skip\footins}{7mm}
% \footnoterule{1cm}
\renewcommand\footnoterule{\vspace*{2mm}\hrule width 1.3cm\vspace*{2mm}}

\title{\Large \bf Dark Matter and Baryogenesis from Visible-Sector Long-Lived Particles} 

\author{Rouzbeh Allahverdi$^{1}$}
\email{rouzbeh@unm.edu}
\author{Ngo Phuc Duc Loc$^{1}$}
\email{locngo148@gmail.com}
\author{Jacek K.~Osi{\'n}ski$^{2}$}
\email{jaksaosinski@gmail.com}
\affiliation{$^{1}$~Department of Physics and Astronomy, University of New Mexico, Albuquerque, NM 87131, USA}
\affiliation{$^{2}$~Astrocent, Nicolaus Copernicus Astronomical Center of the Polish Academy of Sciences, ul.~Rektorska 4, 00-614 Warsaw, Poland}

%%%%%%%%%%%%%%%%%%%%%%%%%%%%%%%%%%%%%%%%%%%%%%%%%%%%%%%%%%%
\begin{abstract}

We present a minimal extension of the standard model that includes a long-lived fermion with weak-scale mass and an ${\cal O}({\rm GeV})$ fermionic dark matter candidate both of which are coupled to quarks. Decays of a TeV-scale colored scalar in a radiation-dominated phase bring the former to a thermal abundance while also producing dark matter. The long-lived fermion then dominates the energy density of the Universe and drives a period of early matter domination. It decays to reheat the Universe, mainly through baryon-number-violating interactions that also generate a baryon asymmetry, with a small branching fraction to dark matter. 
%The dark matter relic abundance also receives a, %typically subdominant, contribution from   
%The dark matter relic abundance receives contributions %from decays of the long-lived fermion and the colored %scalar. 
We find the allowed parameter space of the model and show that it 
%that yields the observed dark matter relic abundance and %baryon asymmetry, 
can be probed by proposed long-lived particle searches as well as next-generation neutron-antineutron oscillation experiments. This model provides a robust explanation of dark matter and baryogenesis as long as the Universe is in a radiation-dominated phase at $T \gtrsim {\cal O}({\rm TeV})$.

%show that a nonstandard cosmological history with a %period of early matter domination driven by a sub-TeV %visible-sector particle can arise rather naturally. This 
%scenario involves a long-lived standard model singlet %that acquires a thermal abundance at high temperatures %from decays and inverse decays of a parent particle with %SM charge(s), and subsequently dominates the energy %density of the Universe as a frozen species. Entropy %generation at the end of early matter domination dilutes %the abundance of dangerous relics (such as gravitinos) %by a factor as large as $10^4$. 
%The scenario can accommodate the correct dark matter %relic abundance for cases with $\langle \sigma_{\rm ann} %v \rangle_{\rm f} \lessgtr 3 \times %10^{-26}$cm$^3$s$^{-1}$. 
%More importantly, the allowed parameter space can be %directly probed by proposed searches for neutral long-%lived particles at the energy frontier, allowing us to %use particle physics experiments to reconstruct the %cosmological history just prior to big bang %nucleosynthesis. 

\end{abstract}
\maketitle

\vspace{-1.1cm}
%%%%%%%%%%%%%%%%%%%%%%%%%%%%%%%%%%%%%%%%%%%%%%%%%%%%%%%%%%%%
\section{Introduction}
\vspace{-1mm}

There are various lines of evidence pointing to the existence of dark matter (DM) in the Universe~\cite{BHS}. However, the identity of DM remains a profound problem at the interface of cosmology and particle physics. In addition, explaining the observed DM abundance depends on the details of the thermal history of the early Universe. Thermal freeze-out in a radiation-dominated (RD) Universe can yield the correct relic abundance for a specific value of the (thermally averaged) annihilation rate $\langle \sigma_{\rm ann} v \rangle = 3 \times 10^{-26}$cm$^3$s$^{-1}$.
%at the time of freeze-out, when \(T = T_{\rm f}\). 
On the other hand, in nonstandard thermal histories this can be achieved for both much larger or smaller values of $\langle \sigma_{\rm ann} v \rangle$
%if the Universe is not in a RD phase at the time of %freeze-out
~\cite{KT}. Important classes of early-Universe models (notably those arising from string theory) typically lead to nonstandard thermal histories that involve one or more epochs of early matter domination (EMD)
%, which is a generic feature of early Universe models %from string theory constructions
~\cite{KSW}. 
%provides an important such example. 
An EMD phase is driven by a matter-like species that comes to dominate the energy density of the Universe at early times and subsequently decays to establish RD prior to big bang nucleosynthesis (BBN). 
%Various production mechanisms during EMD can yield the %correct DM abundance for both $\langle \sigma_{\rm ann} %v \rangle_{\rm f} \lessgtr 3 \times %10^{-26}$cm$^3$s$^{-1}$. An epoch of EMD is driven by a %sufficiently long-lived component with the same equation %of state as matter. 
%can lead to an EMD phase in the postinflationary %Universe if it constitutes a sizeable fraction of the %total energy density and is sufficiently long lived. 
%In the context of string theory, 
This can be due to coherent oscillations of a scalar field like a string modulus displaced from the minimum of its potential during inflation,
%that are displaced from the minimum of their potential %during inflation. Modulus fields have long lifetimes due %to their gravitationally suppressed couplings to other %fields. An EMD era may also be driven by 
or from long-lived nonrelativistic quanta
%that are 
produced in the postinflationary Universe
%and dominate the energy density before decaying. This %can happen, for example, in models that involve hidden %sectors
~\cite{Ng,Hooper,Scott2,Cirelli}. Scenarios with EMD have novel predictions that can be tested  
%that can be tested via 
%for cosmological 
in observations and indirect DM searches~\cite{E1,E2,E3,E4,E5}. 
%A natural question that arises is whether one could also %directly probe the particle(s) driving an epoch of EMD %in laboratory experiments.      

%Motivated by this, we present a scenario where 
We recently proposed a scenario~\cite{VSEMD} where a visible-sector long-lived particle (LLP) with a weak-scale mass drives an EMD epoch. 
%This scenario involves a standard model (SM) singlet, %$N$, that 
The LLP reaches thermal equilibrium
%in a RD Universe 
at temperatures well above its mass,
%as a result of decays and inverse decays of a parent %particle, $X$, with SM charges. $N$ quanta 
maintains a frozen comoving number density, and dominates the energy density of the Universe. It eventually decays to standard-model (SM) particles
%via an effective interaction mediated by $X$. Decay of %$N$ is a higher-order process that is also suppressed by %three-body phase space and by powers of $m_N/m_X$, which %results in a sufficiently long lifetime 
which can occur all the way up to the onset of BBN.
%, with $\tau_N \sim 0.1$ s. We discuss an explicit %realization of this scenario in a minimal extension of %the SM at TeV scale that can also accommodate DM and %baryogenesis. 
%
%The late decay of $N$ dilutes dangerous relics like %unstable gravitinos, as well as any overabundance of DM %or baryon asymmetry, from earlier stages by a factor as %large as $10^4$. Depending on the relation between $m_N$ %and the DM mass, 
It was shown that this scenario can accommodate the observed DM abundance for both large and small values of $\langle \sigma_{\rm ann} v \rangle$. Moreover, the parameter space of this scenario may be 
%directly probed at the energy frontier, such as 
probed by the proposed LLP searches
%for long-lived particles 
at the Large Hadron Collider (LHC)~\cite{MAT1,MAT2,MAT3}. 

Here, we present an explicit realization of this scenario that yields the correct DM abundance and gives rise to successful baryogenesis. The model is a minimal extension of the SM that involves three new fields with baryon-number-violating couplings to quarks: an ${\cal O}({\rm TeV})$ colored scalar $X$ and two Majorana fermions $N$ (a neutral LLP) and $\chi$ (the DM candidate) with masses $m_N \sim 100$ GeV and $m_\chi \approx {\cal O}({\rm GeV})$ respectively. $N$ acquires a thermal-equilibrium abundance through $X$ decays at $T \gtrsim m_N$ and, due to its long lifetime, dominates the energy density of the Universe and drives an epoch of EMD. It eventually decays to establish a RD Universe before the onset of BBN. The main contributions to the DM relic abundance come from $X$ and $N$ decays, while baryon-number-violating decays of $N$ also generate a baryon asymmetry through one-loop electroweak corrections. 

We find the allowed parameter space of the model where the observed DM abundance and correct baryon asymmetry are obtained for some benchmark points that are within or close to the LHC reach. The allowed region overlaps with the sweet spot of the proposed searches for hadronically decaying LLPs~\cite{MAT1,MAT2,MAT3}. Moreover, the model also has predictions for baryon-number-violating processes at low energies. As we show, its parameter space can be probed by next-generation neutron-antineutron oscillation experiments. A positive signal in these experiments, combined with the requirement to satisfy very stringent bounds from double proton decay, will then help us further constrain
%the flavor structure of 
the baryon-number-violating couplings of the LLP in this model. This model is robust as it is largely independent of the details of the postinflationary history provided that the Universe is in a RD phase at $T \gtrsim {\cal O}({\rm TeV})$.

The rest of this paper is organized as follows. In Section~\ref{sec:model}, we present the model and obtain the resulting DM relic abundance and baryon asymmetry.  
%our scenario and the
%conditions that need to be satisfied for its success. We %also present a minimal extension of the SM that can %explicitly realize the scenario. 
In Section~\ref{sec:results}, we present our main results and identify the allowed region of the parameter space of the model. 
%We focus on the $m_N-\tau_N$ plane 
In Section~\ref{sec:signals}, we discuss experimental signals of the model
%at LLP searches and low-energy processes 
within this allowed region focusing on LLP searches and low-energy processes. 
%and determine regions that yield the correct DM %abundance. We also discuss the prospects for probing the %$m_N-\tau_N$ plane in tandem with the long-lived %particle searches at the LHC. 
We conclude the paper in Section~\ref{sec:concl}. Some details of our calculations are presented in the Appendix.  

\section{The Model}\label{sec:model}
% \vspace{-1.2mm}

%We now present a specific realization of the scenario %introduced above. It 
The model is based on a minimal extension of the SM that was proposed for low-scale baryogenesis and DM~\cite{RB} (see~\cite{Rabi} for a supersymmetric version). Using two-component Weyl fermions, the Lagrangian is:
\begin{eqnarray} \label{lagran}
{\cal L} \supset (h_{i} X N u^c_{i} + h^{\prime}_{i j} X^* {d^c_i} d^c_j + h^{\prime \prime}_{i} X \chi u^c_{i} + {1 \over 2} m_N N N + {1 \over 2} m_\chi \chi \chi + {\rm h.c.}) + 
m^2_X |X|^2 ,
\end{eqnarray}
where $u^c$ and $d^c$ denote the left-handed up-type and down-type antiquarks respectively, and $h^\prime_{i j}$ is an antisymmetric tensor. Flavor indices are denoted by $i$ and $j$, while color indices are omitted for simplicity. $X$ is an iso-singlet color-triplet scalar of hypercharge $+4/3$. $N$ and $\chi$ are singlet fermions\footnote{One may forbid $N$ and $\chi$ coupling to leptons by invoking a $Z_2$ symmetry or a continuous symmetry (like lepton number).}
%be charged under a higher-ranked gauge group that %includes the SM.} 
with $m_\chi \ll m_N \ll m_X$. As discussed in~\cite{RB}, $\chi$ is absolutely stable if $m_p - m_e \leq m_\chi \leq m_p + m_e$. It is therefore a DM candidate in this mass window and, in addition, can help address the baryon-DM coincidence puzzle.

We denote the maximum value of $\vert h_i \vert, \vert h^{\prime}_{ij} \vert, \vert h^{\prime \prime}_i \vert$ by $h, h^{\prime}, h^{\prime \prime}$ respectively, and quarks and antiquarks by \(q\) and \(\bar{q}\). Then, the (rest frame) decay width of $X$ is given by:
%
% \begin{equation} \label{Xdec}
% \Gamma_X = \Gamma_{X \rightarrow N} + \Gamma_{X \rightarrow \chi} + \Gamma_{X \rightarrow d^c_j d^c_k} \simeq {(h^2 + {h^{\prime \prime}}^2 + 2 {h^{\prime}}^2) \over 16 \pi} m_X , 
% \end{equation}
%
\begin{equation} \label{Xdec}
\Gamma_X = \Gamma_{X \rightarrow N} + \Gamma_{X \rightarrow \chi} + \Gamma_{X \rightarrow {\bar q} {\bar q}} \simeq {(h^2 + {h^{\prime \prime}}^2 + 2 {h^{\prime}}^2) \over 16 \pi} m_X , 
\end{equation}
where the factor of 2 in front of ${h^{\prime}}^2$ accounts for the two color combinations in $\bar{q}\bar{q}$. The rate for $N$ self-annihilation at energies $E \ll m_X$ follows:
%
% \begin{equation} \label{Nself}
% \Gamma_{\rm self} = \Gamma_{N N \rightarrow u_i u^c_i} \simeq 3 ~ {h^4 \over 16 \pi} {E^2 \over m^4_X} n_N .    
% \end{equation}
% 
\begin{equation} \label{Nself}
\Gamma_{\rm self} = \Gamma_{N N \rightarrow q {\bar q}} \simeq 3 ~ {h^4 \over 16 \pi} {E^2 \over m^4_X} n_N ,    
\end{equation}
where \(n_N\) is the number density of \(N\). The factor of 3 accounts for the color of $q$ in the final state. 
% We omit flavor indices in the subscript of the rate for readability, however we note that the above relation corresponds to the particular flavor combination that leads to the largest value of the coupling. 
$N$ can also annihilate with quarks, which results in:
%
% \begin{equation} \label{Nann}
% \Gamma_{\rm ann} = \Gamma_{N u_i \rightarrow d^c_j d^c_k} + \Gamma_{N d_j \rightarrow u^c_i d^c_k} + \Gamma_{N d_k \rightarrow u^c_i d^c_j} \simeq {3 \times 6} ~ {h^2 {h^{\prime}}^2 \over 16 \pi} {E^2 \over m^4_X} n_{u,d} ,
% \end{equation}
%
\begin{equation} \label{Nann}
\Gamma_{\rm ann} = \Gamma_{N q \rightarrow {\bar q} {\bar q}} \simeq {3 \times 6} ~ {h^2 {h^{\prime}}^2 \over 16 \pi} {E^2 \over m^4_X} n_q ,
\end{equation}
where \(n_q\) is the quark number density, and the factor of 6 is due to the possible color combinations in $q \bar{q} \bar{q}$.
%, and 3 takes $N$ annihilation with $u$ and either of $d$ quarks into account. 
The rest-frame width for three-body decay of $N$ is:
%
% \begin{equation} \label{Ndec}
% \Gamma^{3-{\rm body}}_N = \Gamma_{N \rightarrow u^c_i d^c_j d^c_k} + \Gamma_{N \rightarrow u_i d_j d_k} + \Gamma_{N \rightarrow u_i u^c_i \chi} \simeq {2 \times 6} ~ {h^2 ({h^{\prime}}^2 + {h^{\prime \prime}}^2) \over 128 \cdot 192 \pi^3} {m^5_N \over m^4_X} .
% \end{equation}
%
\begin{equation} \label{Ndec}
\Gamma^{3-{\rm body}}_N = \Gamma_{N \rightarrow {\bar q} {\bar q} {\bar q}} + \Gamma_{N \rightarrow q q q} + \Gamma_{N \rightarrow q {\bar q} \chi} \simeq {2 \times 6} ~ {h^2 ({h^{\prime}}^2 + {h^{\prime \prime}}^2) \over 128 \cdot 192 \pi^3} {m^5_N \over m^4_X} .
\end{equation}
%
%where the factor of 2 arises because $N$ is a Majorana %fermion and decays to both $u d d$ and $u^c d^c d^c$ %final states. 
Note that this decay violates baryon number and, as we will describe below, can be the origin of baryon asymmetry in the model. There is also two-body radiative decay of $N$ at the one-loop level. The corresponding decay width, after all multiplicity factors for the particles in the loop are taken into account, is given by~\cite{ADG}:  
\begin{equation}\label{Ndec2}
\Gamma^{2-{\rm body}}_N = \Gamma_{N \rightarrow \chi \gamma} \simeq {\alpha_{\rm em} h^2 {h^{\prime\prime}}^2 \over 32 \pi^4} {m^3_N \over m^2_X} ,     
\end{equation}
where $\alpha_{\rm em}$ is the electromagnetic fine structure constant. The total decay width of $N$ is $\Gamma_N = \Gamma^{2-{\rm body}}_N + \Gamma^{3-{\rm body}}_N$.   

\vskip 2mm
This model leads to the following thermal history, starting with a RD phase at an initial temperature $T_{\rm i} \gtrsim m_X$, with \(H\) being the Hubble rate: 
%that is established at the end of inflationary reheating %(for reviews, see~\cite{ABCM,Aminreview}) or from the decay %of a heavy modulus, 
% this model leads to the following thermal history. 
%the important stages of the thermal history in our %scenario, arranged by decreasing Hubble rate, are as %follows:
%
\vskip 2mm
\noindent
{\bf (1)} $H \gtrsim H(T = m_X)$. Initially, $X$ is in thermal equilibrium due to its gauge interactions with the SM particles. Its decay and inverse decay to $N$ and $\chi$ then results in (for details, see Appendix~\ref{app:A}):
\begin{eqnarray}
&& n_N = F(\gamma_N) n^{\rm eq}_N \, , \\
&& n_\chi = F(\gamma_\chi) n^{\rm eq}_\chi \, ,
\end{eqnarray}
where $F$ is given in Eq.~(\ref{F}) and \(n^{\rm eq}_{N,\chi}\) denotes the equilibrium value of the \(N\) or \(\chi\) number density.
%
%$N$ also acquires a thermal abundance, $n_N \propto T^3$ %and $\rho_N \propto T^4$, provided that the partial decay %width of $X$ to $N$ satisfies:
%
%\begin{equation} \label{equi}
%\Gamma_{X \rightarrow N} \gtrsim H(T = m_X) . 
%\end{equation}
%
%This ensures that $X$ decay and inverse decay reach %equilibrium before $X$ becomes nonrelativistic (see %Appendix A for details).
%
\vskip 2mm
\noindent
{\bf (2)} $H(T = m_X) > H \gtrsim H_{\rm dom}$. The Universe remains in a RD phase during this stage. $N$ particles are relativistic as long as $H \gtrsim H(T = m_N)$, and their number density follows $n_N \propto a^{-3}$ (where $a$ is the scale factor of the Universe)\footnote{More precisely, $g_* n_N \propto a^{-3}$ as long as $N$ is in chemical equilibrium with the thermal bath, while $n_N \propto a^{-3}$ when it is chemically decoupled. The difference is negligible though as the number of relativistic degrees of freedom, $g_*$, changes minimally for $m_N \lesssim T < m_X$.}.
%
%\vskip 2mm
%\noindent
%{\bf (3)} 
At $H \sim H(T = m_N)$, $N$ quanta become nonrelativistic and their comoving energy density becomes frozen, remaining constant for $H \gtrsim H_{\rm dom}$ provided that: 
%the rate for $N N \rightarrow \psi \psi$ and $N \psi %\rightarrow \psi \psi$ satisfy: 
%
\begin{eqnarray}
&& \Gamma_{\rm self} < H(T = m_N) \label{selfann} \, , \\
&& \Gamma_{\rm ann} < H(T= m_N) \label{ann} \, ,
\end{eqnarray}
which ensures that $N$ self-annihilation and its annihilation with SM particles are inefficient at $T \lesssim m_N$. $N$ starts to dominate the energy density of the Universe once its energy density becomes comparable to that of radiation.
%$\rho_N \simeq \rho_{\rm r}$, 
This happens at $H = H_{\rm dom}$, which is found to be~\cite{VSEMD}: 
% 
%\begin{eqnarray}
%&& \rho_N \simeq 2 \times {3\zeta(3) \over 4\pi^2} m^4_N ~ %{g_{*{\rm dom}} \over g_{*N}} {T^3_{\rm dom} \over m^3_N} %\, \label{Hdom1} \, , \\
%&& \rho_{\rm R} = {\pi^2 \over 30} g_{*{\rm dom}} ~ %T^4_{\rm dom} \, . \label{Hdom3} 
%\end{eqnarray}
%
%Here, $\rho_N \approx m_N n_N$ for $T < m_N$, and we have %used the fact that $n_N \propto a^{-3} \propto s^3$ during %RD, where $s = (2 \pi^2/45) g_* T^3$ is the comoving %entropy density of the Universe at temperature %$T$\footnote{Recall that $s$ remains constant during RD, %which implies that $g_* T^3 \propto a^{-3}$.}. The number %of relativistic degrees of freedom at $T = m_N$ and $T = %T_{\rm dom}$ are given by $g_{*N}$ and $g_{*{\rm dom}}$ %respectively. The factor of 2 in Eq.~(\ref{Hdom1}) accounts %for the two degrees of freedom associated with the Majorana %fermion $N$. Since the total energy density at $H = H_{\rm %dom}$ is $\rho_N + \rho_{\rm R} \simeq 2 \rho_N$, we have %$T_{\rm dom} \simeq (45/\pi^2 g_{*{\rm dom}})^{1/4} (H_{\rm %dom} M_{\rm P})^{1/2}$. Therefore, after using the relation %$m_N \simeq (90/\pi^2 g_{*N})^{1/4} (H(T=m_N) M_{\rm %P})^{1/2}$, we arrive at:
%
% \begin{equation} \label{Hdom2}
% H_{\rm dom} \simeq \sqrt{2}\left(\frac{45\zeta(3)}{\pi^4}\right)^2 {g^{1/2}_{*{\rm dom}} \over g^{5/2}_{*N}} F^2(\gamma_N) H(T = m_N) , 
% \end{equation}
\begin{equation} \label{Hdom2}
H_{\rm dom} \simeq 0.4 {g^{1/2}_{*{\rm dom}} \over g^{5/2}_{*N}} F^2(\gamma_N) H(T = m_N) , 
\end{equation}
where \(g_{*{\rm dom}}\) denotes the number of relativistic degrees of freedom at \(N\) dominance, and \(g_{*N}\) is taken at \(T = m_N\)\footnote{To compute the number of relativistic degrees of freedom at a given temperature, \(g_*(T)\), we have made use of the smooth function generated from the data presented in Table S2 of \cite{Borsanyi_gstar} with cubic spline interpolation.}. A necessary condition for $N$ dominance is:
\begin{equation} \label{dom}
\Gamma_N < H_{\rm dom},
\end{equation}
otherwise, $N$ will decay before dominating and the Universe will remain in a RD phase.
%where $\Gamma_N$ is the width of $N$ decay into three SM %particles.
%
\vskip 2mm
\noindent
{\bf (3)} $H_{\rm dom} > H \gtrsim \Gamma_N$. The Universe enters an EMD era
%driven by $N$ during this stage, which 
that lasts until $N$ quanta decay, at which point
% $N$ decay reheats 
the Universe is reheated to a temperature $T_{\rm dec}$ given by $T_{\rm dec} = (90/g_{*{\rm dec}} \pi^2)^{1/4} (\Gamma_N M_{\rm P})^{1/2}$. 
%Note that $N$ decay must complete 
This must happen before the onset of BBN
%corresponding to $T_{\rm BBN} \simeq 4-5$ MeV
\footnote{An \(\mathcal{O}(1)\) MeV lower bound on the final reheat temperature is studied in~\cite{Kohri} considering the thermalization process for neutrinos including neutrino oscillations and self interactions for both radiative and hadronic decays. A bound of the same order is obtained in~\cite{CMBBBN} from CMB anisotropies considering detailed production of the relic neutrino background with three-flavor oscillations. }, 
%For our definition of $T_{\rm dec}$, 
which translates to $T_{\rm dec} \gtrsim 7-8$ MeV in our case. 
%which implies that:
%
%\begin{equation} \label{onset}
%\Gamma_N \gtrsim \mathcal{O}(50) ~ {\rm s}^{-1}.
%\end{equation}
% 
The entropy released by $N$ decay 
%releases entropy and 
dilutes any preexisting relic abundance by a factor $d$, which is given by~\cite{VSEMD}: 
%To obtain $d$, we note that in the absence of an EMD %phase $n/s$ remains constant for any species with frozen %comoving number density. However, during the EMD epoch $n %\propto a^{-3} \propto H^{2}$, which implies that:
%
%\begin{eqnarray}
%{n(H \sim \Gamma_N) \over n(H = H_{\rm dom})} \simeq %\left({\Gamma_N \over H_{\rm dom}}\right)^2 . 
%\end{eqnarray} 
%
%The ratio of the entropy density at the end of this epoch %(when $H \sim \Gamma_N$) to that at its beginning (when %$H = H_{\rm dom}$) is:
%
%\begin{equation}
%{s(H \sim \Gamma_N) \over s(H= H_{\rm dom})} \simeq %{g_{*{\rm dec}} \over g_{*{\rm dom}}} {T^3_{\rm dec} %\over T^3_{\rm dom}} ,
%\end{equation}
%
%where $T_{\rm dec}$ is the temperature after $N$ decay %completes. Then, after using the relations $T_{\rm dec} %\simeq (90/\pi^2 g_{*{\rm dec}})^{1/4} (\Gamma_N M_{\rm %P})^{1/2}$ and $T_{\rm dom} \simeq (45/\pi^2 g_{*{\rm %dom}})^{1/4} (H_{\rm dom} M_{\rm P})^{1/2}$, the dilution %factor $d$ (which is the ratio of $n/s$ at $H = H_{\rm %dom}$ to that at $H \sim \Gamma_N$) is found to be:
%
\begin{equation} \label{d3}
d \simeq 2^{3/4} \left({g_{*{\rm dec}} \over g_{*\rm dom}} \right)^{1/4} \left({H_{\rm dom} \over \Gamma_N}\right)^{1/2} = 2 {T_{\rm dom} \over T_{\rm dec}} ,
\end{equation}
where $T_{\rm dom} = (45/g_{*{\rm dom}} \pi^2)^{1/4} (H_{\rm dom} M_{\rm P})^{1/2}$. We see from Eqs.~(\ref{Hdom2}) and (\ref{d3}) that: 
\begin{equation} \label{dilution} 
d \simeq 10^{-2} F(\gamma_N) ~ {m_N \over T_{\rm dec}}.
\end{equation}

\vskip 2mm
The normalized number density of $N$ at the time of its decay, called the ``yield" $Y_N$, is given by:
\begin{equation} \label{Nabun}
Y_N \equiv \left({n_N \over s} \right)_{\rm dec} \approx 0.8 F(\gamma_N) d^{-1} = {3 T_{\rm dec} \over 4 m_N}.    
\end{equation}
%
%where $F$ is given in Eq.~(\ref{F}) and $d$ is dilution %factor due to an epoch of EMD driven by $N$. 
%where the last equality is found from conservation of energy. 
We see 
% from Eqs.~(\ref{Hdom2}) and (\ref{d3}) 
that $F(\gamma_N)$ drops out of the final expression for $Y_N$, which can be understood as a smaller $n_N$ implies a later $N$ dominance and hence a smaller dilution factor\footnote{We note that if the condition in Eq.~(\ref{dom}) is not satisfied, there will be no EMD epoch. In this case, we have $d = 1$ and $Y_N < 3 T_{\rm dec}/4 m_N$.}.

%We note that $F(\gamma_N) \approx 1$ for $\gamma_N %\gtrsim 1$, while $F(\gamma_N) \simeq 3 \gamma_N$ if %$\gamma_N \ll 1$, and $d \geq 1$. The maximum value of %$(n_N/s)_{\rm dec}$ is achieved when $X$ decay (and %inverse decay) brings $n_N$ to its equilibrium value and %there is no EMD phase.

The relic density of DM in our model is given by:
\begin{equation} \label{DMabun1}
{n_\chi \over s} \approx 4\times 10^{-3} F(\gamma_\chi) d^{-1} + {3 \over 4} {\rm Br}_{N \rightarrow \chi}{T_{\rm dec} \over m_N} ,
\end{equation}
where:
%
% \begin{equation} \label{branching}
% {\rm Br}_{N \rightarrow \chi} \equiv {\Gamma_{N \rightarrow u_i u^c_i \chi} + \Gamma_{N \rightarrow \chi \gamma} \over \Gamma_N}.    
% \end{equation}
%
\begin{equation} \label{branching}
{\rm Br}_{N \rightarrow \chi} \equiv {\Gamma_{N \rightarrow q {\bar q} \chi} + \Gamma_{N \rightarrow \chi \gamma} \over \Gamma_N}.    
\end{equation}
The first term on the right-hand (RH) side of Eq.~(\ref{DMabun1}) is the contribution from $X$ decay\footnote{As we have demonstrated in Appendix~\ref{app:B}, this is the main contribution to the $\chi$ abundance from the thermal bath. Additionally, \(N q \leftrightarrow \chi q\) scatterings mediated by \(X\), which could establish chemical equilibrium between \(N\) and \(\chi\), are subdominant at temperatures \(T < m_X\) for the small values of \(h^{\prime\prime}\) presented in Section \ref{sec:results}. }, while the second term accounts for production from $N$ decay. This must match the observed relic abundance that follows:
\begin{equation} \label{obsabun}
\left({n_\chi \over s} \right)_{\rm obs} \sim {4 \times 10^{-10}} \left( {1 ~ {\rm GeV} \over m_\chi} \right) \approx 4 \times 10^{-10} ,     
\end{equation}
where the last relation is due to $m_\chi \approx m_p$. Given that $F(\gamma_N) \leq 1$, we see from Eqs.~(\ref{dilution}) and (\ref{DMabun1}) that:
\begin{equation} \label{DMabun2}
{n_\chi \over s} \lesssim \left(F(\gamma_\chi) + {3 \over 4} {\rm Br}_{N \rightarrow \chi}\right) \left({T_{\rm dec} \over m_N}\right) .
\end{equation}

Matching the observed DM abundance thus requires that $T_{\rm dec} \lesssim 10^{-9} m_N$ if $F(\gamma_\chi) \sim 1$ or ${\rm Br}_{N \rightarrow \chi} \sim 1$. The lower limit of $T_{\rm dec} \gtrsim 7-8$ MeV from BBN then results in $m_N \gtrsim 10^7$ GeV (and hence $m_X \gtrsim 10^7$ GeV). This is far beyond the reach of the LHC. Therefore, to have any hope of testing the model in the proposed LLP searches, we must have $F(\gamma_\chi) \ll 1$ and ${\rm Br}_{N \rightarrow \chi} \ll 1$. These imply that DM production from the thermal bath occurs in the freeze-in regime, and $N$ decay to final states with DM is subdominant.      

The baryon-number-violating interactions of $X$ in Eq.~(\ref{lagran}) allow for this model to generate the observed baryon asymmetry of the Universe. Baryogenesis from two-body decays of $X$ (if $m_X > m_N$) or $N$ (if $m_N > m_X$) has been studied in the non-supersymmteric~\cite{RB,ADD} and sypersymmetric versions of the model~\cite{Rabi,ADMS,ADS,Baryo,Visible}. The latter scenario cannot work in our case as we have taken $N$ to be lighter than $X$. Also, out-of-equilibrium decay of $X$ will yield too little asymmetry because of Boltzmann suppression of $n_X$ at $T < m_X$ and subsequent dilution by the EMD phase.

However, as shown in~\cite{Rabi}, three-body decays of $N$
can also be the source of baryogenesis. The baryon asymmetry of the Universe thus produced is given by:
\begin{equation} \label{BAU1}
\eta_{B} \equiv {n_{B} - n_{\bar B} \over s} = {\Gamma_{N \rightarrow {\bar q} {\bar q} {\bar q}} + \Gamma_{N \rightarrow q q q} \over \Gamma_N} \epsilon_B Y_N .    
\end{equation}
The asymmetry parameter \(\epsilon_B\) is determined by one-loop electroweak corrections giving rise to the following typical leading term~\cite{Rabi}:
\begin{equation} \label{BAU2}
\epsilon_B \sim {\alpha_2 \over 4} {m_c m_s m_t m_b \over m^2_W m^2_N} .      
\end{equation}
Here $\alpha_2$ is the $SU(2)_{\rm W}$ gauge fine structure constant, $m_W$ is the mass of the $W$ boson, and subscripts in the numerator denote quarks.

%At energies $E \ll m_X$, one finds an effective %four-fermion interaction $N u^c_i d^c_j d^c_k$ after %integrating out $X$. %This results in $N$ decay to three %quark, and since $N$ is a %Majorana fermion, three %antiquark final states. 

%$N$ decay can be the origin of baryon asymmetry in the %model %as described in~\cite{Rabi}. 

%Moreover, this model can explain the DM content of the %Universe. In the supersymmetric version, the scalar %partner %of $N$ is a natural DM %candidate~\cite{Rabi,ADMS}. In the %nonsupersymmetric %version, a second copy of $N$ that has %approximately the %same mass as the proton is a viable DM %candidate~\cite{RB}. This can in addition address the %baryon-DM coincidence puzzle.

\vspace{2mm}
%%%%%%%%%%%%%%%%%%%%%%%%%%%%%%%%%%%%%%%%%%%%%%%
\section{Results}\label{sec:results}
\vspace{2mm}

In this section we present our results. We choose two benchmark points $m_X = 3$ TeV with $m_N = 100$ GeV, and $m_X = 1$ TeV with $m_N = 100$ GeV.  
%However, neutral LLPs (like our $N$) with decay lengths %above 100 m are particularly difficult to probe because of %the limited sensitivity of the LHC main detectors. 
%The recently proposed MATHUSLA (MAssive Timing Hodoscope %for Ultra Stable neutraL pArticles) detector %concept~\cite{MAT1} is a minimally instrumented, %large-volume surface detector located near ATLAS or CMS. %It would search for neutral long-live particles (LLPs) %produced in the high luminosity LHC (HL-LHC) collisions.
%, extending the lifetime range by a few orders of %magnitude compared to the main detectors. 
%It could discover LLPs 
%with decay lengths up to $3 \times 10^7$ m, which %correspond %to 
%with lifetimes close to the age of the Universe at the %onset of BBN~\cite{MAT2} such as $N$ in our model.      
%Our benchmark point
This overlaps with the most important physics target of the MATHUSLA (MAssive Timing Hodoscope for Ultra Stable neutraL pArticles) detector concept~\cite{MAT1}, namely hadronically decaying LLPs with masses in the ${\cal O}(10 ~ {\rm GeV}) - {\cal O}(100 ~ {\rm GeV})$ range~\cite{MAT3} (more on this later). It also puts $X$ within or close to the LHC reach.

\begin{figure}[ht!]
    \centering
    \includegraphics[width=0.49\textwidth, trim = .5cm .1cm 1cm .3cm, clip = true]{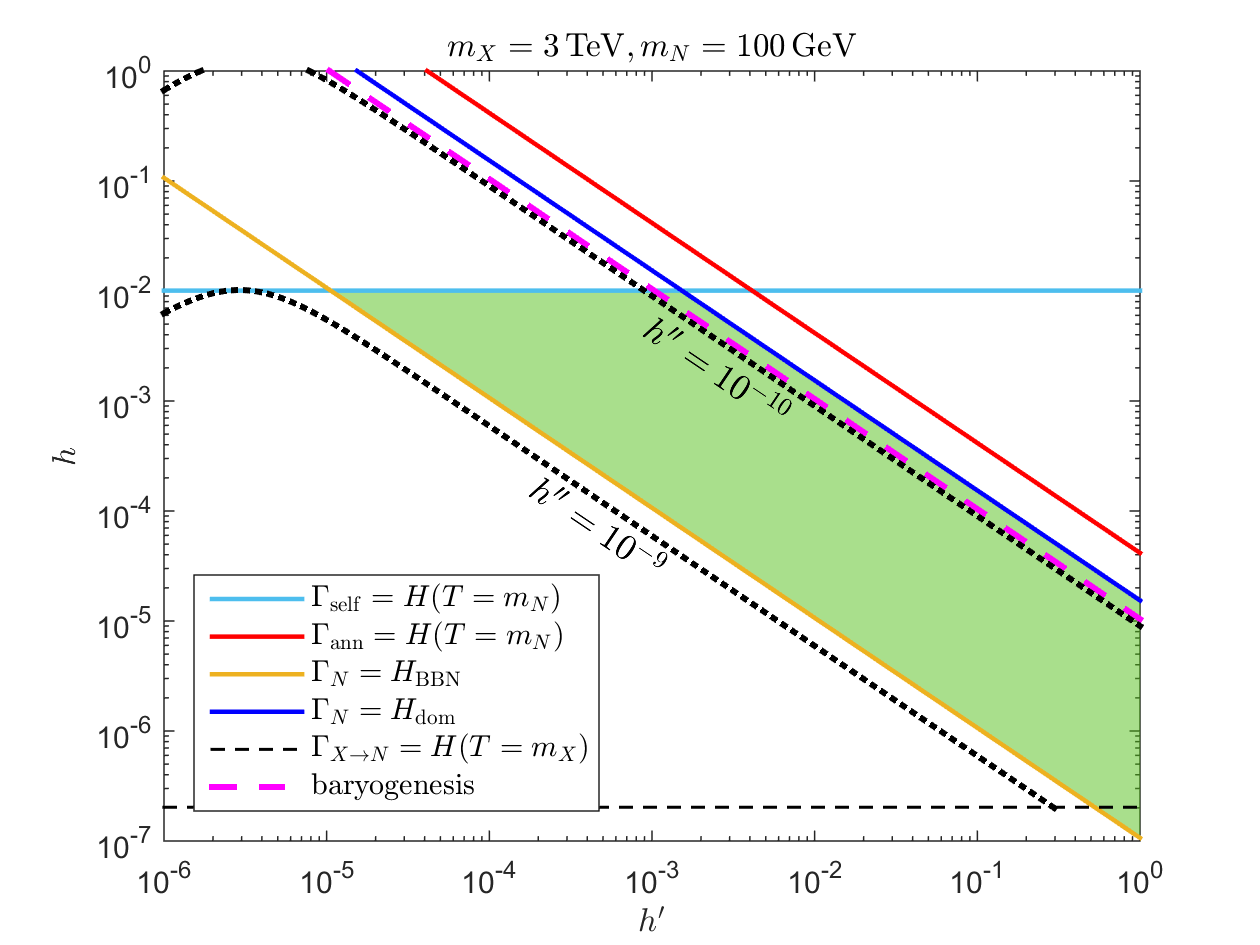}
    \includegraphics[width=0.49\textwidth, trim = .5cm .1cm 1cm .3cm, clip = true]{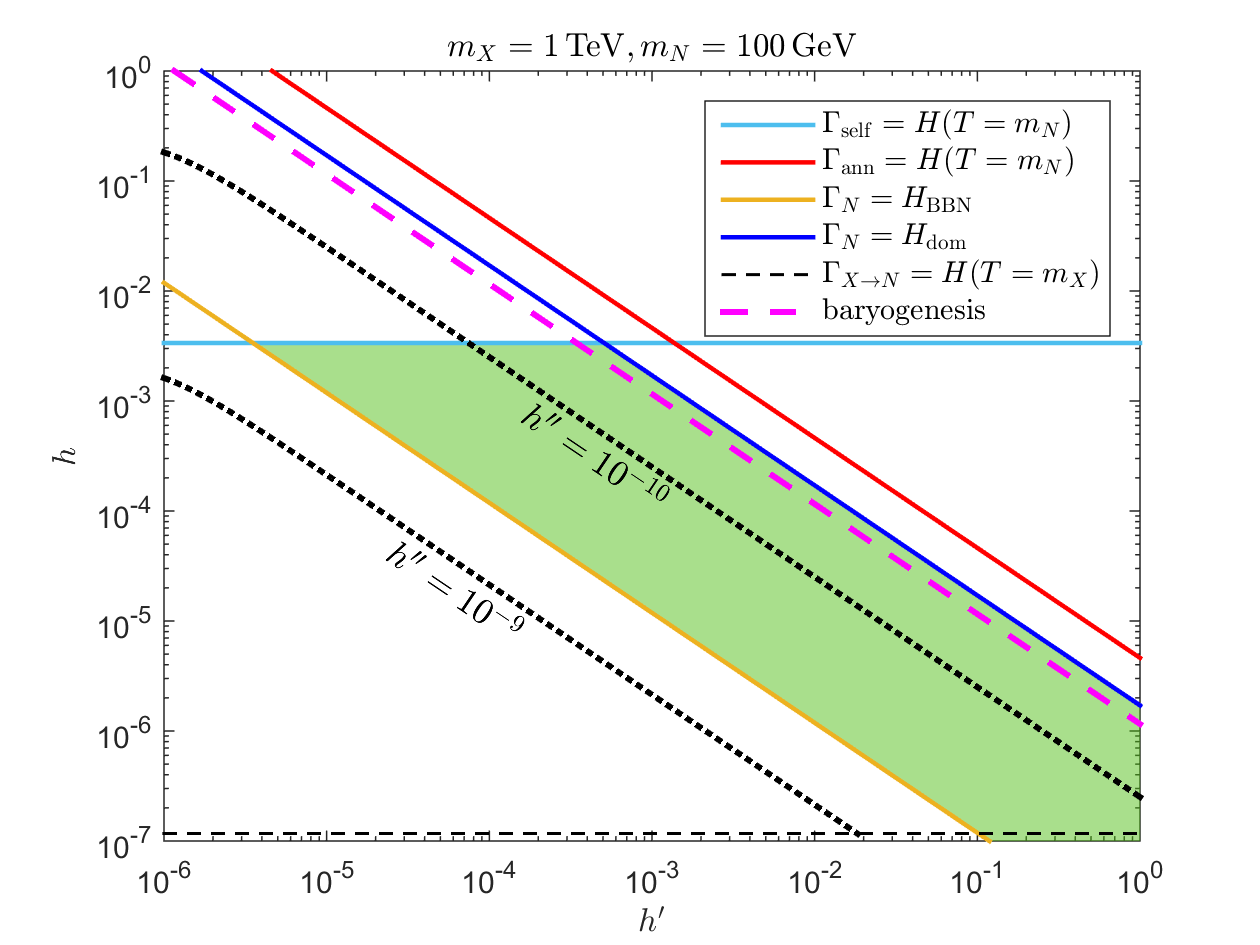}
    \caption{\normalsize Allowed region (shaded green) in the \(h-h^{\prime}\) plane corresponding to EMD driven by \(N\) for our benchmark masses \(m_X = 3\) TeV and \(m_N = 100\) GeV (left), and \(m_X = 1\) TeV and \(m_N = 100\) GeV (right). Dotted black contours indicate values of \(h^{\prime\prime}\) for which the observed DM abundance is obtained. Successful baryogenesis is achieved on the dashed magenta line. 
    % \jo{check if DM contour in the non-EMD region between red and blue lines is calculated without assuming domination}
    }
    \label{fig:h_hprime1}
\end{figure}

In Fig.~\ref{fig:h_hprime1}, we show the region (shaded green) that corresponds to EMD driven by \(N\) in the $h-h^{\prime}$ plane for our benchmark points $m_X = 3$ TeV, $m_N = 100$ GeV 
(left panel) and \(m_X = 1\) TeV, \(m_N = 100\) GeV (right panel). 
% that are in the ballpark for testability of the scenario at colliders.
%\footnote{In passing, we note that in the entire allowed %regions of the right panels of Fig.~\ref{fig:h_hprime}, and %in large parts of the regions in the left panels, the %inequality \(\vert h h^{\prime} \vert < 10^{-4}\), mentioned %for the explicit model in the previous section, is %satisfied.}. 
Below the boundary of the shaded region given by the solid yellow line, decay of $N$ occurs after the onset of BBN. Above the solid blue boundary line, $N$ decays before it has had a chance to dominate the energy density of the Universe. 
% \footnote{In this figure, $F(\gamma_N) = 1$, which implies that $N$ abundance reaches its equilibrium values. For much smaller values of $h$, outside the green band and hence not shown here, we have $F(\gamma_N) < 1$. The blue line has a different slope in this case.}. 
Furthermore, above the solid light-blue line, $N$ self annihilation is efficient such that its abundance is significantly depleted. We additionally show a solid red line above which annihilations of $N$ with the SM bath become efficient, thus the abundance of \(N\) is not depleted in the white band between the solid blue and red lines. 
Above the dashed black line, we have $\Gamma_{X \rightarrow N} \gg H(T = m_X)$, implying that $F(\gamma_N) = 1$ in Eq.~(\ref{Hdom2}) throughout the bulk of the green region. 
% apart from the very small portion just below the horizontal dotted line in which equilibrium is not established. 
Below this line, $F(\gamma_N) < 1$ and \(N\) does not reach an equilibrium abundance. 
% We see that the conditions in Eqs.~(\ref{ann},\ref{dom}), corresponding to the red and blue lines marked 2 and 1 respectively, lie relatively close to each other. This can be understood upon substitution of Eqs.~(\ref{Hdom2},\ref{Nann},\ref{Ndec}) as these conditions have exactly the same dependence on $h$, $h^{\prime}$, $m_X$, and $m_N$ up to an overall numerical factor.

% \vspace{4mm}
The dotted black curves depict contours of \(h^{\prime\prime}\) that achieve the observed DM relic abundance. These contours have two main behaviors depending on the source of the dominant contribution to the DM abundance: \(h \propto 1/h^{\prime}\) when thermal production from \(X\) decay in the bath is dominant, while \(h \propto h^{\prime}\) when direct production from decay of \(N\) is dominant\footnote{For simplicity we have omitted the portion of the \(h^{\prime\prime}\) contours in the region where \(N\) remains out of equilibrium, as it is essentially beyond the reach of our region of interest for our benchmark masses. However the \(h^{\prime\prime}\) contours would change slope and continue as \(h \propto h^{\prime}\) for such small values of \(h\). The slope of the blue boundary line behaves in a similar way.}. We see that the thermal contribution is dominant in the majority of the parameter space shown in the figure. Note that DM is produced in the freeze-in regime as we have $F(\gamma_\chi) \ll 1$ as long as \(h^{\prime\prime} \ll 10^{-7}\) for \(m_X \sim\) TeV. Additionally, for the contribution from \(N\) decay, two-body decay dominates over three-body decay by approximately a factor of \(0.15\,(m_X/m_N)^2\), independent of the couplings as can be seen from Eqs.~\eqref{Ndec} and \eqref{Ndec2}. 
% \jo{mention two slopes of Hdom line?}
% \jo{limit on largest h'' for DM production from X to be out of equilibrium}

We see that successful baryogenesis can be accomplished in the shaded parameter space for our benchmark masses. Due to the smallness of $h^{\prime \prime}$, $N$ dominantly decays to three-body final states $q q q$ and ${\bar q} {\bar q} {\bar q}$ throughout the allowed parameter space. For the chosen value of $m_N = 100$ GeV, Eq.~(\ref{BAU2}) gives $\epsilon_B \sim 3 \times 10^{-8}$ and, in fact, the maximum value of $\epsilon_B$ is obtained for $m_N \sim 100$ GeV\footnote{While this is obvious for larger values of $m_N$, it may seem that smaller values of $m_N$ result in a larger $\epsilon_B$. However, given that quarks in the loop diagram yielding the asymmetry must be on shell, $m_t$ in the numerator of the expression in Eq.~(\ref{BAU2}) will be replaced by $m_u$, which is more than enough to lower $\epsilon_B$ even for $m_N \sim m_b$.}. The dashed magenta line in Fig.~\ref{fig:h_hprime1} corresponds to the observed value of the baryon asymmetry \(\eta_B \sim 10^{-10}\) and it sits within the green EMD region close to the blue boundary which corresponds to moderate dilution. For the small values of \(h^{\prime\prime}\) in the figure, this line scales with \(m_X\) in the same way as the blue boundary line, however \(\epsilon_B\) introduces additional dependence on \(m_N\). Thus for higher values of \(m_X\) and \(m_N\) the baryogenesis line would be pushed outside of the shaded region toward the top-right. In the region between the red and blue lines, where the \(N\) abundance is not depleted by annihilations but there is no period of EMD, the observed DM abundance can be obtained for \(h^{\prime\prime} \approx 6\times 10^{-11}\). However, successful baryogenesis would require \(h^{\prime} \approx 25 h^{\prime\prime}\) which would therefore overproduce DM in this region. 
% \jo{baryon asymmetry line (when independent of \(h^{\prime\prime}\)) scales with \(m_X\) in the same way as the equilibrium \(H_{\rm dom}\) line, but has the extra \(1/m_N^2\) from \(\epsilon_B\) }
% \jo{mention what happens between red and blue lines for values of \(h''\) that give DM and baryogenesis. -- baryogenesis needs \(h' \approx 25 h''\), while observed DM abundance needs \(h'' \approx 6\times 10^{-11}\), so DM would be overproduced for baryogenesis to work in this region}

The effect of very energetic DM particles on the matter power spectrum can further constrain the parameter space. If DM particles are highly relativistic at the time of kinetic decoupling from the thermal bath, their free streaming can prevent the formation of structure at small scales such as those observed in the Lyman-$\alpha$ forest and Milky Way satellite galaxies. As shown in Appendix~\ref{app:C}, this effect does not lead to additional constraints on the parameter space for our benchmark points, but can become important for heavier masses.  

\section{Experimental Signals of the Model}\label{sec:signals}

In this section, we discuss the prospects of probing this model via particle physics experiments.
%other phenomenological and cosmological signatures of %the model. 
%on the parameter space of this model. 
%The baryon-number-violating couplings of $X$ result in  %$n-{\bar n}$ oscillation. It also has novel monojet/monotop %signals, as well as dijet events, at the %LHC~\cite{LHC1,LHC2,ADD}. 
%
\vskip 2mm
\noindent
{\it LLP searches -} The long lifetime of $N$ provides us with an opportunity to directly probe it via the proposed LLP searches at the LHC. The decay length $l_N$ of $N$ particles produced from $X$ decay in colliders is related to the rest-frame lifetime $\tau_N$ via $l_N = {\bar b} c \tau_N$, where ${\bar b}$ is the average boost factor of $N$. 
% We show the $l_N = {\rm const}$ contours in Fig. . 
We find that $l_N > 100$ m within the allowed parameter space, which makes $N$ a natural target for dedicated LLP searches. In particular, MATHUSLA would search for neutral LLPs produced in the high luminosity LHC (HL-LHC) collisions, 
%extending the lifetime range by a few orders of %magnitude compared to the main detectors. It could 
and could discover LLPs with decay lengths up to $3 \times10^7$ m (corresponding to lifetimes close to the age of the Universe at the onset of BBN)~\cite{MAT2}. We note that throughout the allowed parameter space of Fig.~\ref{fig:h_hprime1}, three-body final states $q q q$ and ${\bar q} {\bar q} {\bar q}$ totally dominate $N$ decay. 
%to three-body final states $q q q$ and ${\bar q} {\bar q}
% {\bar q}$ throughout the allowed parameter space.Within %the allowed parameter space of Fig.~\ref{fig:h_hprime1}, 
The lifetime of $N$ in this region varies between $10^{-6}$ s (blue boundary) and $3 \times 10^{-2}$ s (yellow boundary)\footnote{The lifetime of $X$ ranges from \(10^{-18}\) s to \(10^{-26}\) s in our allowed region. This is too short to result in displaced vertices.}, and the corresponding decay lengths are: 
\begin{eqnarray}
&& 5 \times 10^3 ~ {\rm m} \lesssim l_N \lesssim 1 \times 10^8 ~ {\rm m}  ~ ~ ~ ~ ~ (m_X = 3 ~ {\rm TeV})\,, \\ \nonumber  
&& 2 \times 10^3 ~ {\rm m} \lesssim l_N \lesssim 5 \times 10^7 ~ {\rm m}  ~ ~ ~ ~ ~ (m_X = 1 ~ {\rm TeV})\,.
\end{eqnarray}
%
%approximately ${\rm few} \times 10^3$ m to ${\rm few} %\times 10^8$ m. 
Thus, the parameter space of our model overlaps very well with the most important physics target of MATHUSLA (i.e., hadronically decaying LLPs with masses in the ${\cal O}(10 ~ {\rm GeV}) - {\cal O}(100 ~ {\rm GeV})$ range~\cite{MAT3})\footnote{The collaboration has studied the discovery potential of LLPs produced in exotic Higgs decays~\cite{MAT2}, or via mixing with the Higgs~\cite{MAT3}. $X$ decay, as the main channel to produce $N$ in our model, should be implemented properly to calculate the cross-section for $N$ production.}.  
%\jo{For green region with benchmark masses: $N$ lifetime %ranges between \(10^{-6}\) s (blue) to few times \
%(10^{-2}\) s (yellow), with corresponding boosted decay %lengths being approximately few times \(10^3\) m to \
%(10^8\) m; X lifetimes range from \(10^{-20}\) s to \
%(10^{-26}\) s.}
%
%Decay lengths above $10^{-2}$ cm are within the reach of %the LHC. For $10^{-2} ~ {\rm cm} < l_N < 10^2$ cm, $N$ %production and decay gives rise to displaced vertices at %the LHC, while the case with $10^2 ~ {\rm cm} < l_N < %10^4$ cm leads to displaced jet/lepton signals. 
\vskip 2mm
\noindent
{\it Low-energy processes -} It is also possible to probe this model via low-energy experiments. The most stringent experimental bound on the model parameters comes from double proton decay $p p \rightarrow K^+ K^+$.
%\footnote{There are also bounds from $K^0_s$-${\bar %K}^0_s$ and $B^0_s$-${\bar B}^0_s$ mixing, but these are %weaker than 
%easily satisfied for the values satisfying 
%double proton decay limits.}. 
This process is due to the $\Delta B = 2$ and $\Delta s = 2$ effective interaction $(u d s)^2$. The $N$-mediated and $\chi$-mediated contributions to this operator are given by $(h_1 h^{\prime}_{12})^2/16 \pi^2 m^4_X m_N$ and $(h_1 h^{\prime \prime}_{12})^2/16 \pi^2 m^4_X m_\chi$ respectively~\cite{DM}, where we have included flavor indices.
%that arise at low energies
%~\cite{DM}. 
The current experimental limit of $\tau_{p p \rightarrow K^+ K^+} > 1.7 \times 10^{32}$ yr~\cite{double1,double2}, is then translated into an upper bound on $h_1 h^{\prime}_{12}$ for given values of $m_X$ and $m_N$.
Scaling the limit in~\cite{DM} for our benchmark points results in:
\begin{eqnarray} \label{doubleproton}
&& h_1 h^{\prime}_{12} \lesssim 3 \times 10^{-6} ~ ~ ~ , ~ ~ ~ h_1 h^{\prime \prime}_{12} \lesssim 3 \times 10^{-7} ~ ~ ~ ~ ~ (m_X = 3 ~ {\rm TeV}) \, , \\ \nonumber  
&& h_1 h^{\prime}_{12} \lesssim 3 \times 10^{-7} ~ ~ ~ , ~ ~ ~ h_1 h^{\prime \prime}_{12} \lesssim 3 \times 10^{-8} ~ ~ ~ ~ ~ (m_X = 1 ~ {\rm TeV}) \, .
\end{eqnarray}
%
%Both of these constraints are comfortably satisfied in the allowed parameter space shown in Fig. 
%\footnote{There are also bounds from $K^0_s$-${\bar %K}^0_s$ and $B^0_s$-${\bar B}^0_s$ mixing. But those are %easily satisfied for the values that satisfy double %proton decay limits.}. 
%With $h_1 h^{\prime}_{12}$ and $h_1 h^{\prime %\prime}_{12}$ constrained by double proton decay, 
%
Another important limit is from $n-{\bar n}$ oscillation, which can be used to constrain $h_1 {h^{\prime}_{13}}^2$ and $h_1 {h^{\prime \prime}_{13}}^2$. The $\Delta B = 2$ effective operator $(u d b)^2$ gives rise to $n-{\bar n}$ oscillation at the one-loop level~\cite{DM}. The contributions from $N$ and $\chi$ couplings to the oscillation amplitude $G_{n-{\bar n}}$ can be parameterized as $h^2_1 {h^{\prime}_{13}}^4 m_N/16 \pi^2 m^6_X$ and $h^2_1 {h^{\prime \prime}_{13}}^4 m_\chi/16 \pi^2 m^6_X$ respectively (times logarithmic factors 
${\rm ln}(m^2_X/m^2_N)$ and ${\rm ln}(m^2_X/m^2_\chi)$ respectively)~\cite{ADD}. 
The oscillation lifetime is given by $\tau_{n-{\bar n}} \sim (\Lambda^6_{\rm QCD} G_{n-{\bar n}})^{-1}$. The current experimental limit is $\tau_{n-{\bar n}} \geq 3\times 10^8$ s~\cite{nnbar1,nnbar2,nnbar3}, while the next generation experiments predict a sensitivity $\tau_{n-{\bar n}} \geq 5 \times 10^{10}$ s~\cite{nnbar4}.
For our benchmark points, and choosing $\Lambda_{\rm QCD} = 250$ MeV, we find the following range that is compatible with current limits and also within the reach of future experiments:
\begin{eqnarray} \label{nnbar}
&& 3 \times 10^{-6} \lesssim h_1 {h^{\prime}_{13}}^2 \lesssim 4 \times 10^{-5} ~ ~ ~ , ~ ~ ~ 2 \times 10^{-5} \lesssim h_1 {h^{\prime \prime}_{13}}^2 \lesssim 3 \times 10^{-4} ~ ~ ~ ~ ~ (m_X = 3 ~ {\rm TeV}) \, , \\ \nonumber  
&& 1 \times 10^{-7} \lesssim h_1 {h^{\prime}_{13}}^2 \lesssim 2 \times 10^{-6} ~ ~ ~ , ~ ~ ~ 8 \times 10^{-7} \lesssim h_1 {h^{\prime \prime}_{13}}^2 \lesssim 1 \times 10^{-5} ~ ~ ~ ~ ~ (m_X = 1 ~ {\rm TeV}) \, .
\end{eqnarray}
%

\begin{figure}[ht!]
    \centering
    \includegraphics[width=0.49\textwidth, trim = .5cm .1cm 1cm .3cm, clip = true]{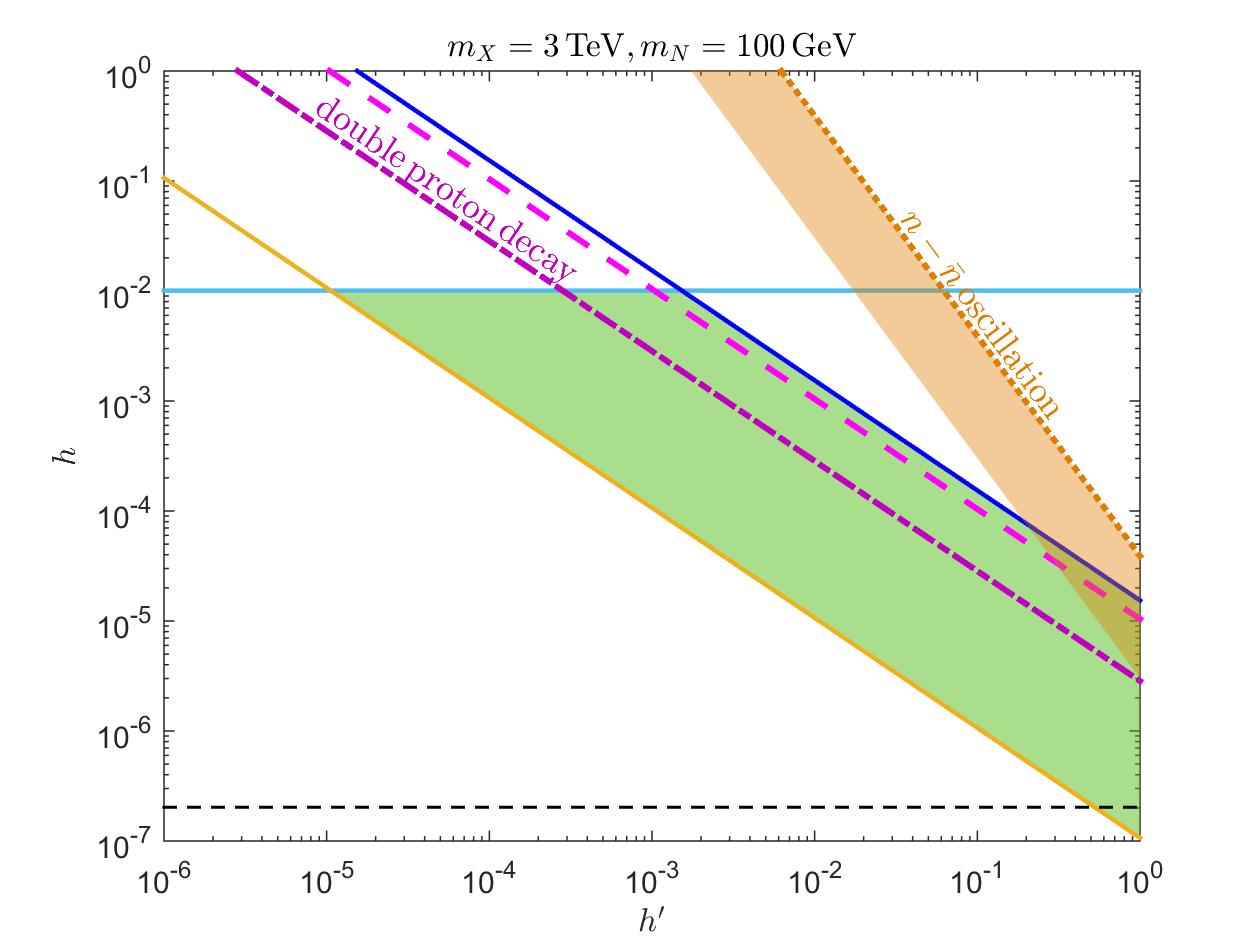}
    \includegraphics[width=0.49\textwidth, trim = .5cm .1cm 1cm .3cm, clip = true]{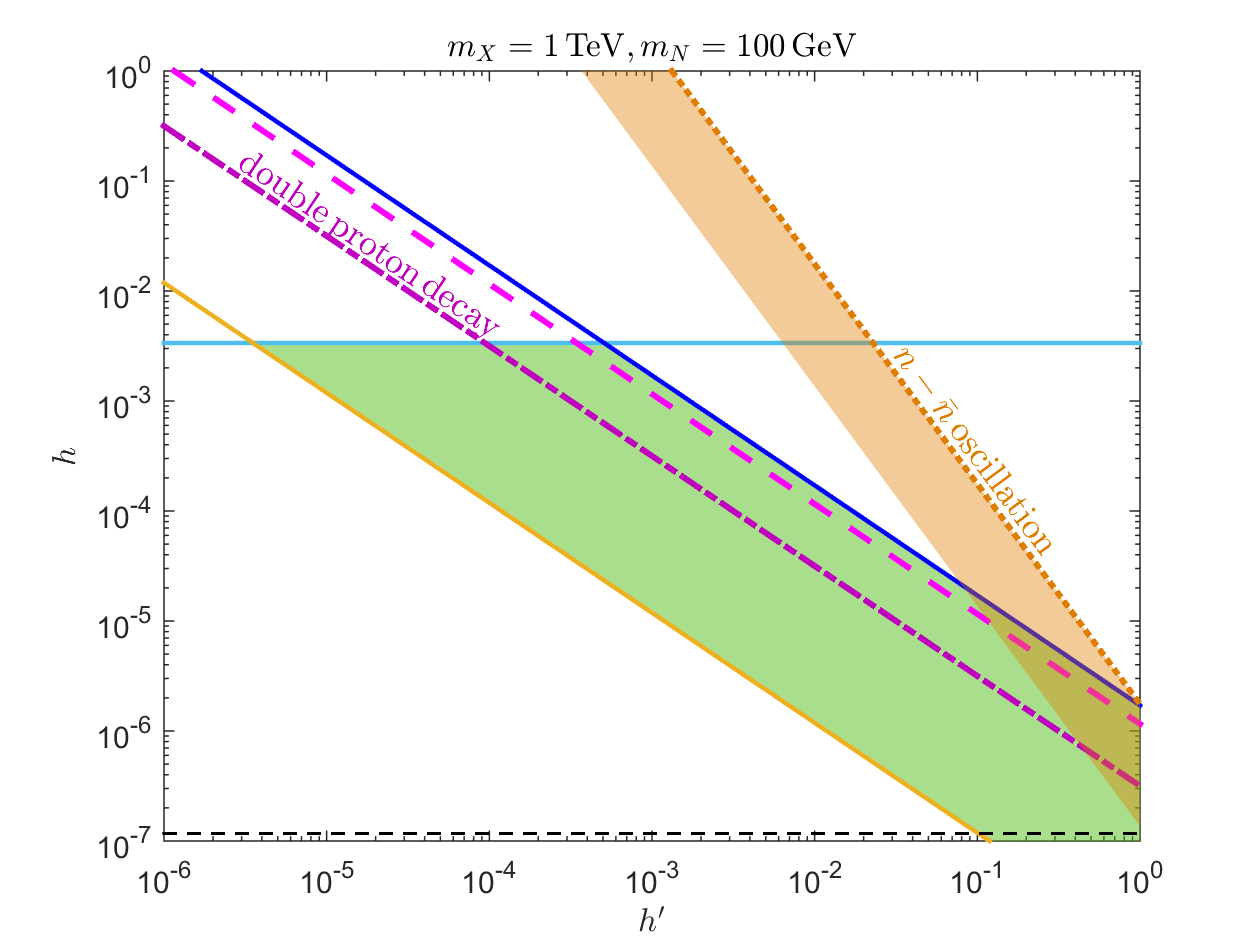}
    \caption{\normalsize The same region as in Fig.~\ref{fig:h_hprime1} but including flavor-dependent constraints from double proton decay (violet dash dot line) and current neutron-antineutron oscillation (orange dotted line). Future sensitivity in neutron-antineutron oscillation experiments is shown by the orange band. The dashed magenta baryogenesis line corresponds to \(h h^\prime \approx 1 \times 10^{-5}\) (left) and \(h h^\prime \approx 1 \times 10^{-6}\) (right). 
    %for our benchmark points $m_X = 3$ TeV, $m_N = 100$ %GeV (left) and $m_X = 1$ TeV (right). 
    %The bounds cut in to the green region from the top %right for small masses. The width of the neutron %band is given by \(\Lambda_{\rm QCD} = 0.2-0.3\) GeV.
    % Dashed contours are DM abundance for \(h'' = %10^{-8,-9,-10}\), thick dash-dot is baryon %asymmetry. Free-streaming bound added to bottom %panel (\(T_{\rm R} \gtrsim 45\) MeV). Gray triangles %show diproton limit for given masses. 
    %\jo{mention that both bounds come in further as \
    %(m_N\) decreases, but then we lose baryogenesis. } 
    }
    \label{fig:h_hprime3}
\end{figure}

%\jo{fix the numbers} \jo{corresponding bounds for \%(\chi\) and \(hh''\) are very weak (larger than 1 for our %values of \(h''\)) for both proton and neutron limits }
\vskip 2mm
We note that the values of $h^{\prime \prime}$ (which is the largest of $h^{\prime \prime}_{ij}$) in Fig.~\ref{fig:h_hprime1} are so small that the upper limit on $h_1 h^{\prime \prime} _{12}$ in Eq.~(\ref{doubleproton}) is comfortably satisfied and $h_1 {h^{\prime \prime}_{13}}^2$ is well outside the range in Eq.~(\ref{nnbar}). In Fig.~\ref{fig:h_hprime3} we show the flavor-dependent bounds from Eqs.~(\ref{doubleproton},\ref{nnbar}) on $h h^{\prime}$ and $h {h^{\prime}}^2$. 
%However, since these bounds are flavor-dependent, they %must be interpreted with care.
%for our benchmark points $m_X = 3$ TeV, $m_N = 100$ GeV 
%(left panel) and $m_X = 1$ TeV, $m_N = 100$ GeV (right %panel). 
In the left-panel, corresponding to $m_X = 3$ TeV and $m_N = 100$ GeV, the double proton bound (violet dash-dot line) is 
%significantly weaker than the upper limit on $h %h^{\prime}$ 
inside the green region, and the $n-{\bar n}$ oscillation band (shaded orange) from Eq.~(\ref{nnbar}) overlaps with a portion of the green region. In the right panel, where $m_X = 1$ TeV and $m_N = 100$ GeV, the situation is similar with slightly more overlap between the orange and green bands.  
%the double proton decay bound still sits well above the %green region. However, the $n-{\bar n}$ oscillation band %is now entirely inside the green band. 

Given that $h$ and $h^{\prime}$ represent the maximum values of $h_i$ and $h^{\prime}_{ij}$, respectively, let us see how the intersection of these limits and the green region 
% orange and green bands 
should be interpreted. We specifically focus on the baryogenesis line that is inside this region where the model produces the observed DM relic abundance and the desired baryon asymmetry. In both panels, the baryogenesis line
%corresponds to \(h h^\prime \approx 1 \times 10^{-5}\) (\%(h h^\prime \approx 1 \times 10^{-6}\)) which 
is above the double proton decay bound and yet consistent with it. We only need \(h_1 h^{\prime}_{12}\) to not be the largest combination of these couplings on this line, i.e., that $h_1$ and/or $h^{\prime}_{12}$ are sufficiently smaller than the largest values of $h_i$ and $h^{\prime}_{ij}$ respectively. As for the $n-{\bar n}$ oscillation limit, any $\tau_{n-{\bar n}} = {\rm const.}$ contour in the orange band is compatible with all points in the green band that are above it. Again, it will be enough if $h_1$ and/or $h^{\prime}_{13}$ are sufficiently smaller than the largest values of the $h_i$ and $h^{\prime}_{ij}$ couplings respectively. On the portion of the baryogenesis line that is within the orange band, both $h_1$ and $h^{\prime}_{13}$ can take their largest values while remaining compatible with the current bound. This results in $4 \times 10^9 ~ {\rm s} \leq \tau_{n-{\bar n}} \leq 5 \times 10^{10}$ s (in the left panel) and $7 \times 10^8 ~ {\rm s} \leq \tau_{n-{\bar n}} \leq 5 \times 10^{10}\,\text{s}$ (in the right panel) for this portion of the baryogenesis line. 
%the combination \(h_1 {h^{\prime}_{13}}^2\) can be the %largest considering 
%This portion satisfies the current $n-{\bar n}$ %oscillation bound and can also be probed in the next %generation experiments. 
A suitable flavor structure in agreement with both experimental bounds (and within the reach of the next generation $n-{\bar n}$ oscillation experiments) is $h^{\prime}_{12},~ h^{\prime}_{23} \ll h^{\prime}_{13}$ and $h_2, ~ h_3 \ll h_1$\footnote{This will also ensure the bounds from the $K_L-K_S$ mass difference as well as $B^0_d$-${\bar B}^0_d$ and $B^0_s$-${\bar B}^0_s$ oscillations are satisfied~\cite{DM}.}. A positive future signal, along with the requirement for successful baryogenesis, can then be used to pinpoint $h_1$ and $h^{\prime}_{13}$. %This, in combination with the double proton bound, will %help us constrain the flavor structure of $h_i$ and %$h^{\prime}_{ij}$ couplings.   

Finally, we comment on the prospect to experimentally probe DM in our model via direct and indirect searches. Couplings of $\chi$ to the quarks, see Eq.~(\ref{lagran}), result in both spin-independent and spin-dependent interactions between DM and nucleons. The spin-independent elastic scattering cross section is $\propto {h^{\prime \prime}_1}^4/m^{8}_X$~\cite{RB}. 
%For the values of $h^{\prime \prime}$ i
%In the allowed parameter space shown in Fig. , 
The resulting $\sigma_{\rm SI}$ within the allowed parameter space in Fig. 1 is many orders of magnitude below the current bounds from direct detection experiments~\cite{xenon1t}. The spin-dependent cross section is found to be $\propto {h^{\prime \prime}_1}^4/m^4_X$~\cite{RB}. While $\sigma_{\rm SD} \gg \sigma_{\rm SI}$, it is still much smaller than the current limits from direct detection searches~\cite{pico}. Also, the DM annihilation rate at the present time follows $\langle \sigma_{\rm ann} v \rangle \propto {h^{\prime \prime}_1}^4/m^4_X$~\cite{RB}. This is far too small to give rise to any detectable signals in indirect detection searches~\cite{fermi1,fermi2} even after enhancement of the small-scale DM perturbations during EMD is taken into account~\cite{E1,E3,E5}\footnote{We note that the allowed parameter space in Fig.~\ref{fig:h_hprime1} implies a moderate duration for EMD. Therefore, the resulting boost in the number of microhalos will not be very significant in this case.}. 

\section{Conclusion}\label{sec:concl}

In this paper, we presented a minimal extension of the SM that gives rise to the observed DM relic abundance and accommodates successful baryogenesis. This model involves an ${\cal O}({\rm TeV})$ colored scalar $X$ and two Majorana fermions $N$ and $\chi$ that are SM singlets. $N$ is a neutral LLP with $m_N \sim {\cal O}(100)$ GeV and $\chi$ is the DM candidate whose stability requires that $m_\chi \approx m_p$. The correct DM abundance and baryon asymmetry are obtained through an interplay between thermal and nonthermal effects. 

Starting in a RD Universe at $T \gtrsim m_X$, decays of $X$ bring $N$ to a thermal abundance while also producing DM in the freeze-in regime. $N$ subsequently dominates the energy density of the Universe as a result of its long lifetime and drives an epoch of EMD. $N$ eventually decays and establishes a RD Universe before the onset of BBN. $N$ decay is an additional source of DM production and, due to its baryon-number-violating nature, also gives rise to baryogenesis. We found the allowed parameter space where the observed DM relic abundance and baryon asymmetry are obtained while satisfying other phenomenological and cosmological constraints. This region of the parameter space overlaps with the MATHUSLA sweet spot for hadronically decaying LLPs, which warrants a careful study of the discovery prospects of our model. It can also be probed by next-generation $n-{\bar n}$ oscillation experiments, and a future positive signal can provide further information about the flavor structure of the baryon-number-violating couplings of the model. 
%couplings to the best opportunity to experimentally %probe this model.  

The model we presented here is robust and largely independent of the postinflationary thermal history as it only requires that the Universe be in a RD phase at $T \gtrsim {\cal O}({\rm TeV})$. This work can be extended in a number of directions. A natural question on the theoretical side is its embedding in UV-complete models of the early Universe like those arising from string theory.
%Explicit models in string constructions typically %predict epochs of EMD driven by string moduli. 
Our model can work in this set up, which typically predicts epochs of EMD driven by string moduli, as long as the last phase of modulus-driven EMD reheats the Universe to $T \gtrsim m_X$. Models with high-scale supersymmetry breaking (for example, see~\cite{ABCO}) can be a suitable framework in this regard. On the phenomenological side, it would be interesting to study %prospects for discovering $N$ at dedicated LLP searches %(especially MATHUSLA) as well as 
dijet and monojet signals associated with $X$ decay at the LHC (similar to the analysis performed in~\cite{ADD}) as a complementary probe of the model parameter space. We leave a detailed investigation along these lines for future work.         

\vspace{-1mm}
%%%%%%%%%%%%%%%%%%%%%%%%%%%%%%%%%%%%%%%%%%
\section*{Acknowledgements}

The work of R.A. and N.P.D.L. is supported in part by NSF Grant No. PHY-2210367. J.O. is supported by the project AstroCeNT: Particle Astrophysics Science and Technology Centre, carried out within the International Research Agendas programme of the Foundation for Polish Science financed by the European Union under the European Regional Development Fund.

%%%%%%%%%%%%%%
% \section*{
% Appendix
% }
\appendix
\renewcommand{\thesubsection}{\Alph{subsection}}

%\vspace{-1cm}
\subsection{Evolution of the Energy Density}\label{app:A}

Here, we derive the time evolution of the number densities of $N$ and $\chi$ particles produced via the reactions $X \leftrightarrow N + q$ and $X \leftrightarrow \chi + q$ respectively.
%interaction term $h X N \psi$ in Eq.~(\ref{Lnew}). 
There are also scatterings between \(N\) and \(\chi\) mediated by \(X\), however, these are subdominant compared to \(X\) decay (see Appendix \ref{app:B}) so we do not include them in the Boltzmann equations below. 
We work in the limit $m_{q} \ll m_N \ll m_X$, hence, for simplicity we take $m_{q}$, $m_\chi$, and $m_N$ to be zero below. 
%This is a good approximation as long as we are interested in %the evolution of $\rho_N$ over time scales where $T \gg %m_N$.
%
%Let us start with 
The equations that govern the occupation numbers of $N$ and $\chi$, denoted by $f_N$ and $f_\chi$ respectively, are:
\begin{eqnarray} \label{EE1}
&& {df_N({\vec p}) \over dt} = \int{{d^3 p_X \over (2 \pi)^3} {d^3 p_{q} \over (2 \pi)^3} {3 h^2 m^2_X \over 8 E_X E_{q} E_N} (f_X({\vec p}_X) - f_{q}({\vec p}_{q}) f_N({\vec p})) (2 \pi)^4 \delta^{(3)} ({\vec p}_X - {\vec p} - {\vec p}_{q}) \delta(E_X - E_N - E_{q})} \, , \\
&& {df_\chi({\vec p}) \over dt} = \int{{d^3 p_X \over (2 \pi)^3} {d^3 p_{q} \over (2 \pi)^3} {3 {h^{\prime \prime}}^2 m^2_X\over 8 E_X E_{q} E_\chi} (f_X({\vec p}_X) - f_{q}({\vec p}_{q}) f_\chi({\vec p})) (2 \pi)^4 \delta^{(3)} ({\vec p}_X - {\vec p} - {\vec p}_{q}) \delta(E_X - E_\chi - E_{q})} \, .
\end{eqnarray}
Here, $f_X$ and $f_{q}$ are the occupation numbers of $X$ and $q$ respectively. 
%{\cal M} = h m_X$ is the Feynman amplitude for the $X %\leftrightarrow N/\chi + \psi^*$ process. 
The factor of 3 on the RH side of these equations accounts for the fact that $X$ is a color triplet.

As shown in~\cite{VSEMD}, these equations can be solved to find:
\begin{eqnarray} \label{fN}
&& f_N({\vec p}) = f^{\rm eq}_N({\vec p}) \left[1 - {\rm exp} \left(-{3\gamma_N T^2_{\rm i}\over 2{p}^2} \int_{m^2_X/T^2_{\rm i}}^{m^2_X/T^2}{{t}^{1/2} e^{-({t} T_{\rm i}/4 p)} d{t}} \right) \right] \, , \\
&& f_\chi({\vec p}) = f^{\rm eq}_\chi({\vec p}) \left[1 - {\rm exp} \left(-{3\gamma_\chi T^2_{\rm i}\over 2{p}^2} \int_{m^2_X/T^2_{\rm i}}^{m^2_X/T^2}{{t}^{1/2} e^{-({t} T_{\rm i}/4 p)} d{t}} \right) \right] \,,   
\end{eqnarray}
where $\gamma_N \equiv \Gamma_{X \rightarrow N}/H(T=m_X)$ and $\gamma_\chi \equiv \Gamma_{X \rightarrow \chi}/H(T=m_X)$. Here, $T_{\rm i}$ is the initial temperature at which we set $f_N(\vec p) = 0$.   

We can then calculate the comoving number densities of $N$ and $\chi$, $n_N^{\rm co} (t)$ and $n_\chi^{\rm co}(t)$ respectively, as functions of time:
\begin{eqnarray}
&& n_N^{\rm co} (t) = {2a^3(t) \over (2 \pi)^3} \int{f_N({\vec p}) d^3 p} = {a^3(t) \over \pi^2} \int{f_N({p}) p^2 d p} = F(\gamma_N,T) n^{\rm eq, co}_N \, , \\
&& n_\chi^{\rm co} (t) = {2a^3(t) \over (2 \pi)^3} \int{f_\chi({\vec p}) d^3 p} = {a^3(t) \over \pi^2} \int{f_\chi({p}) p^2 d p} = F(\gamma_\chi,T) n^{\rm eq,co}_\chi \, ,
\end{eqnarray}
where the factor of 2 counts the internal degrees of freedom of $N$ and $\chi$ and
\begin{eqnarray} \label{F}
F(\gamma,T) \equiv  {\int{f^{\rm eq}_N(p) \left[1 - {\rm exp} \left(-{3\gamma T^2_{\rm i}\over 2{p}^2} \int_{m^2_X/T^2_{\rm i}}^{m^2_X/T^2}{{t}^{1/2} e^{-({t} T_{\rm i}/4 p)} d{t}} \right) \right] p^2 dp} \over \int{f^{\rm eq}_N(p) p^2 dp}}.
\end{eqnarray}
In this expression, \(\gamma\) without a subscript is understood to be either \(\gamma_N\) or \(\gamma_\chi\) as it is the same for both. 

\begin{figure}[ht!]
    \centering
    \includegraphics[width=0.49\textwidth, trim = .5cm 0cm 1cm .4cm, clip = true]{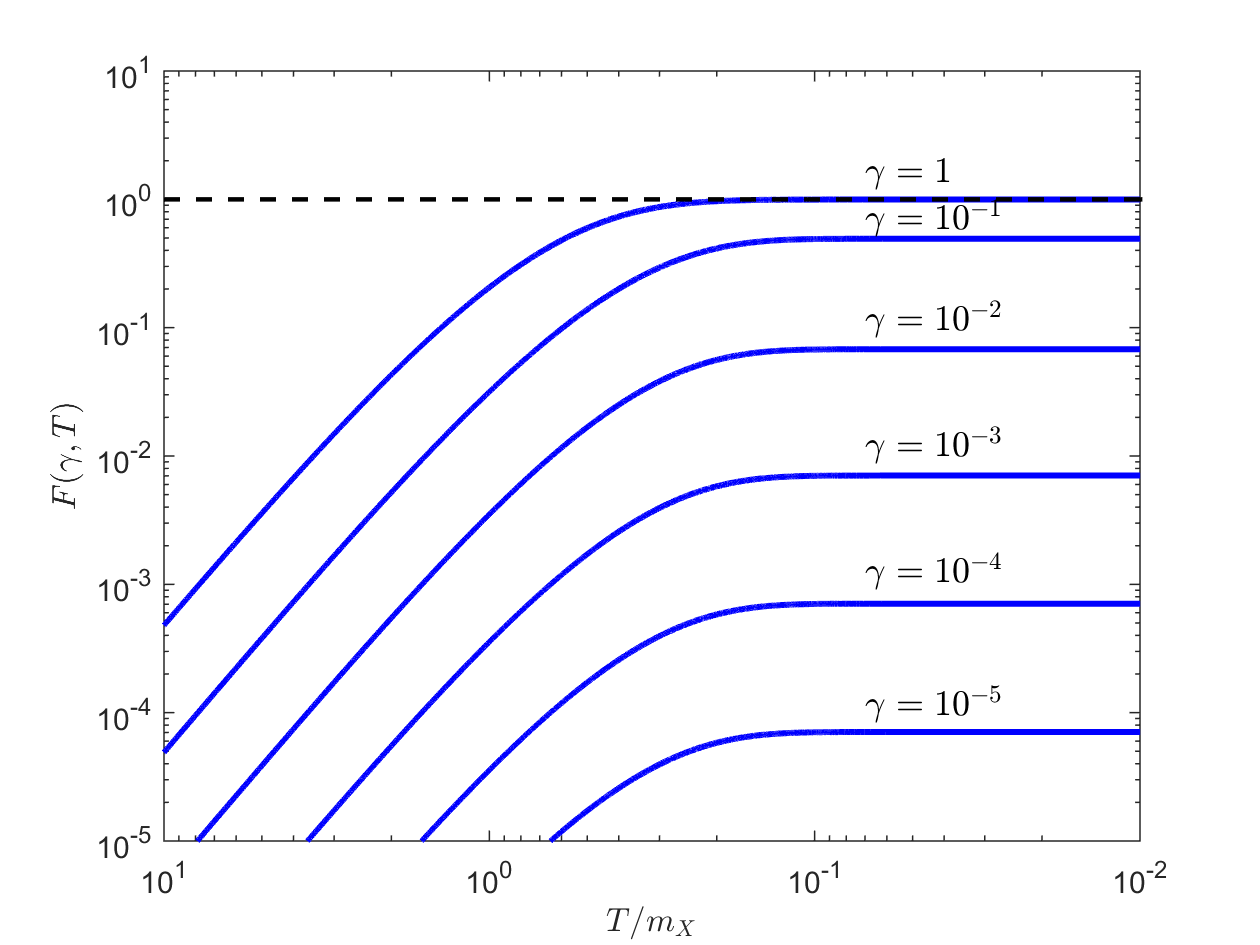}
    \includegraphics[width=0.49\textwidth, trim = .5cm 0cm 1cm .4cm, clip = true]{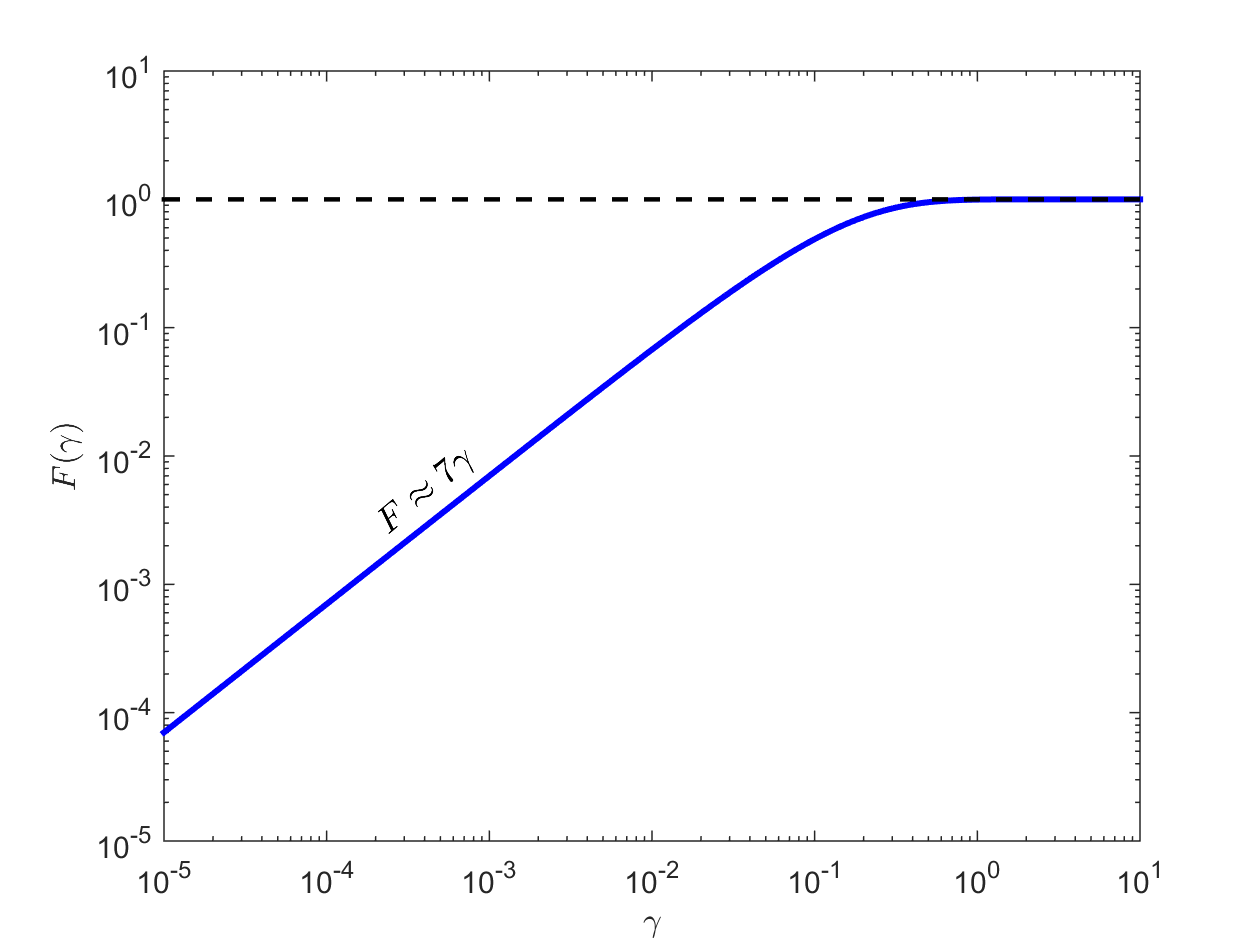}
    \caption{\normalsize 
    % \jo{update caption - mention that gamma applies to both N and chi}
    % \jo{remake with ratio of number densities instead of energy densities, also show \(F(\gamma)\) instead of n/s, add \(\gamma\) labels to lines in left panel} 
    Evolution of the comoving number density of either \(N\) or \(\chi\) produced by \(X\) decay, normalized to the equilibrium number density. The left panel shows evolution with temperature for different values of \(\gamma\) while the right panel shows the dependence of \(F\) on \(\gamma\) at low temperatures (as compared to \(m_X\)) after it has saturated. }
    \label{fig:F(gamma)}
\end{figure}

In the left panel of Fig.~\ref{fig:F(gamma)}, we show \(F(\gamma,T)\) 
% ratios $n^{\rm co}_{N,\chi}/n^{\rm co, eq}_{N,\chi}$ 
as a function of \(T/m_X\). Note that it saturates before \(T \approx 0.1 m_X\) regardless of the value of \(\gamma\). We therefore take \(F(\gamma) = F(\gamma, T \lesssim 0.1 m_X)\) to be the final comoving number density produced by \(X\) decay normalized to the equilibrium value. The right panel of Fig.~\ref{fig:F(gamma)} shows \(F(\gamma)\) as a function of \(\gamma\), and we see that this function is roughly equal to \(F(\gamma) \approx 7\gamma\) in the freeze-in regime (matching the determination in Ref.~\cite{FIMP}), before flattening to \(F(\gamma) \approx 1\) once \(\gamma \gtrsim 1\). Furthermore, there is no dependence on the initial time as long as it is before \(T \sim m_X\). 

%\vspace{-7mm}
%%%%%%%%%%%%%%%%%%%%%%%
\subsection{Contribution of Thermal Bath to DM Relic Abundance}\label{app:B}

Here, we show that the dominant contribution to the DM relic abundance from the thermal bath comes from $X$ decay. As mentioned, we have $F(\gamma_\chi) \ll 1$, and hence DM production from annihilations in the thermal bath occurs in the freeze-in regime. 
%For a constant $\langle \sigma v \rangle$ that does not %depend on $T$, 
$X$ quanta are kinematically accessible to the thermal bath at $T \gtrsim m_X$, and decay of on-shell $X$ quanta (which we discussed in the text) dominates over $X$-mediated annihilations. 
%in the thermal bath. 
At $T_{\rm dom} \lesssim T \lesssim m_X $
%The abundance of DM particles produced from annihilations
%during the RD phase at $T_{\rm dom} \lesssim T \lesssim %m_X$, 
as well as in the adiabatic part of EMD\footnote{Radiation from the prior RD phase dominates over that produced from $N$ decay in this period.
%The DM contribution from the nonadiabatic phase of EMD, %on the other hand, is subdominant in our case.
} 
%follows (see~\cite{AOFI}), we have 
the temperature is simply redshifted according to $T \propto a^{-1}$.
The abundance of DM particles produced from annihilations in this interval
follows (see~\cite{AOFI}):
%\footnote{We note that at $T > m_X$ decay of $X$ %dominates over scatterings in the thermal bath.} 
%and the adiabatic part of the EMD %epoch\footnote{Radiation from the prior RD phase %dominates over that produced from $N$ decay in this %period. The DM contribution from the nonadiabatic phase %of EMD, on the other hand, is subdominant in our case.} %follows (see~\cite{AOFI}):
% \jo{should we mention the nonadiabatic phase?}
%
% \begin{equation} \label{scattabun}
% \left({n_\chi \over s} \right)_{\rm scatt} \simeq {45 \sqrt{90} \zeta(3)^2 \over 2 \pi^7 g^{5/4}_{*X} g^{1/4}_{*{\rm dom}}} \langle \sigma v \rangle M_{\rm P} m_X {T_{\rm dec} \over T_{\rm dom}} .
% \end{equation}
%
\begin{equation} \label{abun1}
\left({n_\chi \over s} \right)_{\rm ann} \sim 10^{-4} \langle \sigma v \rangle (T \sim m_X) M_{\rm P} m_X \left({T_{\rm dec} \over T_{\rm dom}}\right) .
\end{equation}
%
%This happens to be the case for ${\bar q} q \chi \chi$ %and ${\bar l} l \chi \chi$ interaction terms due to %photon exchange at one-loop level where:
%
%\begin{equation} \label{scattcs1}
%\langle \sigma v \rangle \simeq {{h^{\prime \prime}}^{2} %\alpha \over 16 \pi} {1 \over m^2_X} .   
%\end{equation}
%
DM production in the entropy-generating phase of EMD, where $T \propto a^{-3/8}$, mainly occurs at $T \sim m_\chi/4$ resulting in (for example, see~\cite{E1}):
\begin{equation} \label{abun2}
\left({n_\chi \over s} \right)_{\rm ann} \sim 
10^{-2} 
\langle \sigma v \rangle (T \sim m_\chi/4) M_{\rm P} T_{\rm dec} \left({T_{\rm dec} \over m_\chi} \right)^6 .
\end{equation}

The rates for DM production from tree-level $qq/q {\bar q}/{\bar q} {\bar q}$ annihilations (via $X$ exchange) and $q {\bar q}/l {\bar l}$ annihilations at the one-loop level (via photon exchange) respectively follow:
\begin{eqnarray} \label{anns}
&& \langle \sigma v \rangle^{\rm tree}(T) \sim
{{h^{\prime}}^2 {h^{\prime \prime}}^2 + {h^{\prime \prime}}^4 \over 16 \pi} {T^2 \over m^4_X}  
%{T^4_{\rm dec} \over \pi^2 m^4_X} 
\, ,  \\
&& \langle \sigma v \rangle^{1-{\rm loop}}(T) \sim {{\alpha}^2_{\rm em} {h^{\prime \prime}}^4 \over 16 \pi} {1 \over m^2_X} 
%{T^3_{\rm dec} \over \pi^2 m^2_X} 
\, . 
\end{eqnarray}
%
%Here, $\alpha$ denotes the square of the product of %couplings other than $h^{\prime \prime}$ that appear in %the scattering diagram and associated numerical factors. 
%Given that $\alpha \ll 1$, 
Using these expressions, we see that the RH sides of Eqs.~(\ref{abun1},\ref{abun2}) are totally overwhelmed by the first term on the RH side of Eq.~(\ref{DMabun2}) throughout the parameter space shown in Fig.~\ref{fig:h_hprime1}. This confirms that $X$ decay makes the main contribution of the thermal bath to the DM abundance in the allowed parameter space.       
%We also have ${\bar u} u \chi \chi$ and $u d d \chi$ %interaction terms mediated by $X$ for which:
%
%\begin{equation} \label{scattcs2}
%\langle \sigma v \rangle \simeq {{h^{\prime \prime}}^{2} %\alpha \over 16 \pi} {T^2 \over m^4_X} .     
%\end{equation}
%
%Note that this is suppressed $\propto (T/m_X)^2$ %compared with that in Eq.~(\ref{scattcs1}), and hence %these scatterings contribute even less to the DM %abundance.

One may also consider \(N q \leftrightarrow \chi q\) scatterings mediated by \(X\) that could in principle establish chemical equilibrium between \(N\) and \(\chi\). However, for the values of \(h^{\prime\prime}\) in Fig.~\ref{fig:h_hprime1}, the rate for this process is well below the Hubble rate at temperatures \(T < m_X\). This is expected since if \(X\) decay cannot bring \(\chi\) into equilibrium, the scattering process, which has a smaller rate, will be even less efficient. 

%%%%%%%%%%%%%%%%%%%%%%%%
\subsection{Constraints from DM Free Streaming} \label{app:C}

To see the importance of free streaming of DM particles, we first find the rate for scattering of DM particles off the SM particles in the thermal bath at the time of their production from $N$ and $X$ decay. 
%Recall that the main sources of DM production in our %model are direct decay of $N$ and $X$ decay.  
%Another constraint comes from the effect of very energetic %DM particles on the matter power spectrum. DM particles are %highly relativistic at the time of kinetic decoupling from %the thermal bath, their free streaming can prevent the %formation of structure at small scales such as satellite %galaxies and those observed in the Lyman-$\alpha$ forest. 
%The two main sources of DM production in our case are %direct decay of $N$ and $X$ decay. 

$\bullet$ {\bf DM from $N$ decay}. DM particles produced from $N$ decay have a typical momentum $p \simeq m_N/2$ when most production happens at $H \sim \Gamma_N$. 
%(recall that two-body decays dominate in the allowed %parameter space). 
The rates for $\chi$ scattering off $u$ quarks (via $X$ exchange at tree-level) and scattering off quarks and leptons (via photon exchange, which arises at the one-loop level) in the thermal bath respectively follow:  
\begin{eqnarray} \label{rate1}
&& \Gamma^{\rm tree}_{\rm scatt}(T = T_{\rm dec}) \sim {3 {h^{\prime \prime}}^4 \over 16 \pi^3} {m_N T^4_{\rm dec} \over m^4_X} \, ,  \\
&& \Gamma^{1-{\rm loop}}_{\rm scatt}(T = T_{\rm dec}) \sim {\alpha^2_{\rm em} {h^{\prime \prime}}^4 \over 16 \pi^3} {T^3_{\rm dec} \over m^2_X} \, . 
\end{eqnarray}
where $(m_N T_{\rm dec})^{1/2}$ in the numerator of the first expression is the momentum in the center-of-mass frame at $H \sim \Gamma_N$. 
%In addition, $\chi$ scatters off quarks and leptons in the thermal bath via photon exchange arising at one-loop %level. The corresponding scattering rate is given by:
%
%\begin{equation} \label{rate2}
%\Gamma_{\rm scatt}(T = T_{\rm dec}) \sim {{\alpha}^2_{\rm %em} {h^{\prime \prime}}^4} {T^3_{\rm dec} \over \pi^2 %m^2_X}.      
%\end{equation}
%
%Given that ${h^{\prime \prime}}^2 \ll m_X/M_{\rm P}$, %because $F(\gamma_\chi) \ll 1$, we see that both rates are %much smaller than $H_{\rm dec} \sim T^2_{\rm dec}/M_{\rm %P}$ in the allowed parameter space in Fig. . This implies %that highly relativistic DM particles produced from $N$ %decay are kinetically decoupled\footnote{In fact, many %scatterings are needed for DM to achieve and remain in %kinetic equilibrium with the thermal bath (for example, %see~\cite{KMY}).} and freely stream, which could pose %problems for structure formation. 
%of these relativistic DM particles
%, given that $m_N \gg m_\chi$, and undergo free %streaming. 
%This can prevent the formation of structure at small %scales such as satellite galaxies and those observed in %the Lyman-$\alpha$ forest. 
%If most of the DM particles are produced from $N$ decay, %the effect on the matter power spectrum can be used to %constrain the $m_N - T_{\rm dec}$ plane~\cite{E4}. 
%fraction of DM particles with relativistic %velocities~]cite{E4}. This can be translated into a bound on %${\rm Br}_{N \rightarrow \chi}$ in our model. 

%However, the dominant contribution to DM relic abundance %in our case comes from $X$ decay. 
$\bullet$ {\bf $X$ decay.} Another source of DM production is $X$ decay. As shown above, for $F(\gamma_\chi) \ll 1$, the comoving number density of $\chi$ is saturated at $T \gtrsim 0.1 m_X$. We can therefore take the typical momentum of DM particles at this time to be $p \sim m_X/2$. The rates for $\chi$ scattering off $u$ quarks (via $X$ exchange) and scattering off quarks and leptons (via photon exchange) at that time are respectively given by:
\begin{eqnarray} \label{rate2}
&& \Gamma^{\rm tree}_{\rm scatt}(T = 0.1 m_X) \sim {3 {h^{\prime \prime}}^4 \over 16 \pi^3} m_X 
%{T^4_{\rm dec} \over \pi^2 m^4_X} 
\, ,  \\
&& \Gamma^{1-{\rm loop}}_{\rm scatt}(T = 0.1 m_X) \sim {{\alpha}^2_{\rm em} {h^{\prime \prime}}^4 \over 16 \pi^3} m_X 
%{T^3_{\rm dec} \over \pi^2 m^2_X} 
\, . 
\end{eqnarray}

The rate to maintain kinetic equilibrium $\Gamma_{\rm kin}$ is much smaller than $\Gamma_{\rm scatt}$ as many scatterings are needed for DM to exchange enough energy with the thermal bath 
%achieve and remain in kinetic equilibrium with the thermal %bath 
(for example, see~\cite{KMY}). This implies that while $\Gamma_{\rm scatt} > H$ is a necessary condition, it will not be sufficient to ensure that DM particles are in kinetic equilibrium with the thermal bath. On the other hand, DM particles are kinetically decoupled from the thermal bath whenever $\Gamma_{\rm dec} < H$. 

%Since ${h^{\prime \prime}}^2 \ll m_X/M_{\rm P}$, 
%We see that both rates are much smaller than $H(T = 0.1 %m_X)$, Since ${h^{\prime \prime}}^2 \ll m_X/M_{\rm P}$, and %hence DM particles produced from $X$ decay are kinetically %decoupled upon production. The momentum of these $\chi$ %particles is simply redshifted $\propto a^{-1}$ due to the %expansion of the universe. This implies that the typical %momentum of these particles at $H \sim \Gamma_N$ is $\ll %T_{\rm dec}$. 

From Eq.~(\ref{rate1}) we see that $\Gamma_{\rm scatt} (T = T_{\rm dec}) \ll H_{\rm dec}$, where $H_{\rm dec} \sim T^2_{\rm dec}/M_{\rm P}$, throughout the allowed parameter space shown in Fig.~\ref{fig:h_hprime1}. 
% This implies that DM particles produced in $N$ decay are kinetically decoupled from the beginning. 
This implies that DM particles produced from $N$ decay are kinetically decoupled at $H_{\rm dec}$, and hence their typical momentum at $H_{\rm dec}$ is $p \sim m_N/2$.  
%Given that ${h^{\prime \prime}}^2 \ll m_X/M_{\rm P}$, %because $F(\gamma_\chi) \ll 1$, we see that both rates %are much smaller than $H_{\rm dec} \sim T^2_{\rm %dec}/M_{\rm P}$ in the allowed parameter space in Fig. . %This implies that highly relativistic DM particles %produced from $N$ decay are kinetically decoupled and %freely stream, which could pose problems for structure %formation.
Similarly, it turns out from Eq.~(\ref{rate2}) that $\Gamma_{\rm scatt} (T = 0.1 m_X) \ll H(T = 0.1 m_X)$ in the allowed parameter space. Thus, DM particles produced from $X$ decay have an initial momentum $p \sim m_X/2$ that is redshifted $\propto a^{-1}$ due to the expansion of the Universe. As a result, their momentum is $p \ll T_{\rm dec}$ at the end of the EMD epoch when the Hubble rate is $H_{\rm dec}$\footnote{As discussed in \cite{quasidec}, if DM decouples during the entropy-generating phase of EMD, it enters a quasidecoupled state until the end of EDM, increasing the final momentum as compared to a fully decoupled species. The DM momentum can then sit somewhere between the decoupled value \(p \ll T_{\rm dec}\) and the thermal value \(p \sim T_{\rm dec}\). In either case, however, the DM particles remain essentially nonrelativistic in our region of interest, as \(T_{\rm dec} \lesssim \) GeV at the top-right edge of the green EMD region in Fig.~\ref{fig:h_hprime1}.}. 
% \jo{mention quasi-decoupling and why it is not really important in a footnote}   

%In our model, the two main sources of DM production are %$N$ and $X$ decays, and we consider DM kinetic %decoupling in Appendix \ref{} with both sources in mind. 
% 
%The free-streaming of highly-relativistic DM particles %can pose problems for small-scale structure formation as %probed by the Lyman-\(\alpha\) forest and Milky Way %satellite galaxies. 
A detailed analysis in~\cite{E4} has constrained the boost factor of DM particles at the end of EMD as a function of $T_{\rm dec}$ based on their impact on the matter power spectrum. 
%If most of the DM particles were produced from $N$ %decay, this would restrict $T_{\rm dec}$ as a function %of $m_N$, as the DM mass is known in our case. 
For fixed values of $m_N$ and $m_X$, this translates into a bound in the $h-h^{\prime}$ plane. For the benchmark points discussed in Section~\ref{sec:results}, $X$ decay is the dominant source of DM production within the allowed parameter space, and hence the typical boost factor of DM particles at $H_{\rm dec}$ is small. As a result, 
the constraint from the suppression of the matter power spectrum is easily satisfied throughout the entire parameter space shown in Fig.~\ref{fig:h_hprime1}. 

\begin{figure}[ht!]
    \centering
    \includegraphics[width=0.49\textwidth, trim = .5cm .1cm 1cm .3cm, clip = true]{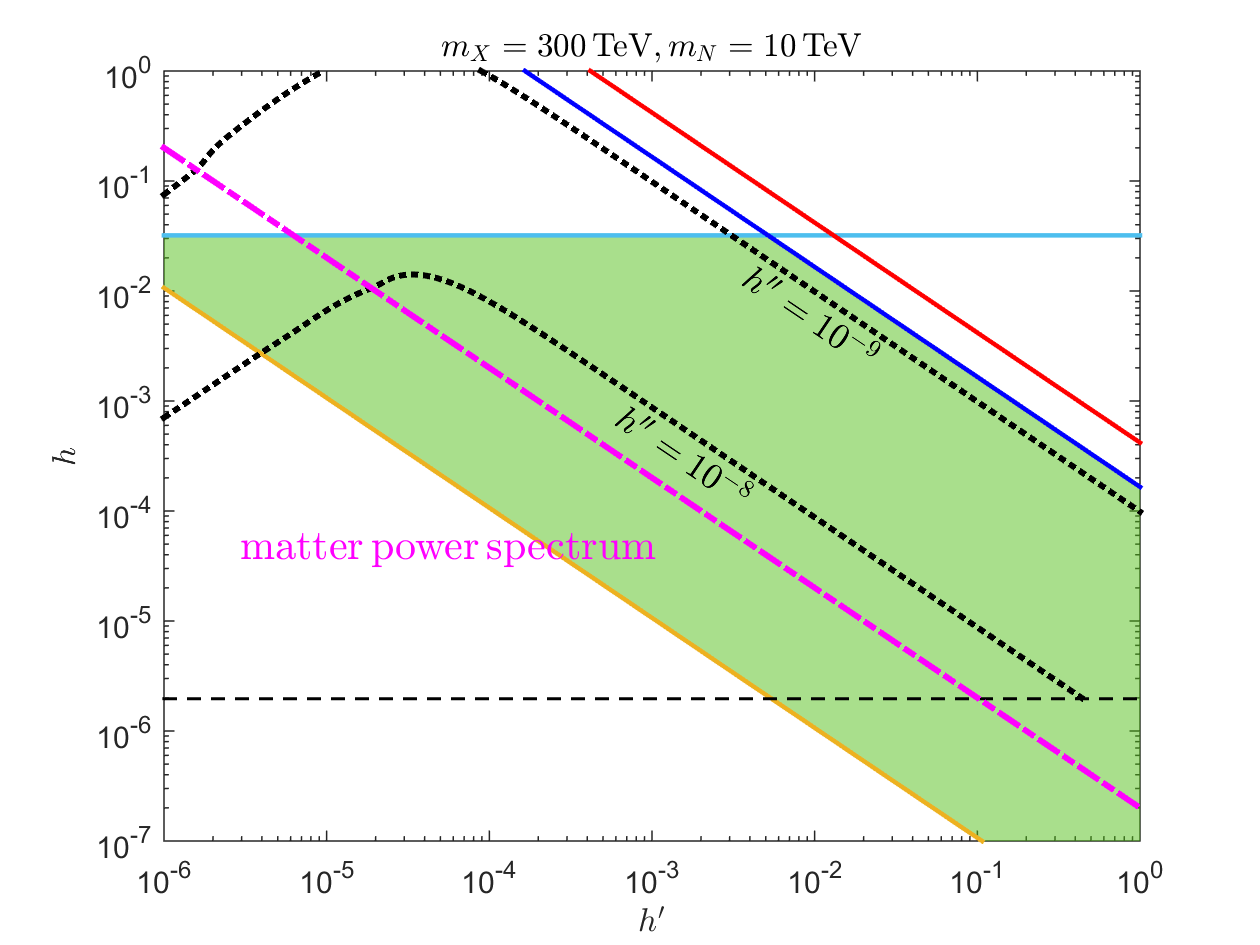}
    \caption{\normalsize Allowed region (shaded green) in the \(h-h^{\prime}\) plane for the high-mass case \(m_X = 300\) TeV and \(m_N = 10\) TeV. Constraints on the matter power spectrum from structure formation (magenta dash-dot line) cut in to the green region from the bottom left for such high masses, which is particularly relevant in the top-left corner where the DM abundance originates primarily from \(N\) decay. 
    % \jo{check if DM contour in the non-EMD region between red and blue lines is calculated without assuming domination}
    }
    \label{fig:h_hprime2}
\end{figure}

For heavier masses, however, limits from free-streaming may become relevant.  
%as discussed in the Appendix.
% However, the situation changes if $N$ decay is the dominant source of DM production, leading to much larger boost factors for the bulk of the DM particles. 
In Fig.~\ref{fig:h_hprime2} we show the \(h-h^{\prime}\) plane for a high-mass case with \(m_X = 300\) TeV and \(m_N = 10\) TeV, 
% Though the DM abundance is still dominated by \(X\) decay throughout the majority of the shaded parameter space, \(N\) decay dominates in the top-left corner where the dotted black \(h^{\prime\prime}\) contours turn over. 
as well as a dash-dot magenta line that corresponds to the \(2\sigma\) confidence bound from~\cite{E4} on the scale of suppression of the matter power spectrum, which is related to the free-streaming length. We see that in this high-mass case, our allowed parameter space is partially constrained from the bottom-left by these considerations. This is particularly relevant in the top-left corner where DM is produced primarily from \(N\) decay as can be seen from the turnover in the \(h^{\prime\prime}\) contours. In the rest of the green region, most of the DM particles are produced from $X$ decay and have a much smaller boost factor ($\sim T_{\rm dec}/m_\chi$ instead of $\sim m_N/m_\chi$). 

% \newpage
%%%%%%%%%%%%%%


\begin{thebibliography}{99}


\bibitem{BHS}
G. Bertone, D. Hooper, and J. Silk, Phys. Rept. {\bf 405}, 279 (2005).


\bibitem{KT}
M. Kamionkowski and M. S. Turner, Phys. Rev. D {\bf 42}, 3310 (1990).


\bibitem{KSW}
For a recent review, see: G. Kane, K. Sinha, and S. Watson, Int. J. Mod. Phys. D {\bf 24}, 1530022 (2015).


% \bibitem{Howie}
% H. Baer, K-Y Choi, J. E. Kim, and L. Roszkowski, Phys. Rept. {\bf 555}, 1 (2015).


\bibitem{Ng}
J. A. Dror, E. Kuflik, and W. H. Ng, Phys. Rev. Lett. {\bf 117}, 211801 (2016).


\bibitem{Hooper}
A. Berlin, D. Hooper, and G. Krnjaic, Phys. Rev. D {\bf 94}, 095019 (2016).


\bibitem{Scott2}
J. A. Dror, E. Kuflik, B. Melcher, and S. Watson, Phys. Rev. D {\bf 97}, 063524 (2018).


\bibitem{Cirelli}
M. Cirelli, Y. Gouttenoire, K. Petraki, and F. Sala, JCAP {\bf 1902}, 014 (2019).


\bibitem{E1}
A. L. Erickcek, Phys. Rev. D {\bf 92}, 103505 (2015).


\bibitem{E2}
A. L. Erickcek, K. Sinha, and S. Watson, Phys. Rev. D {\bf 94}, 063502 (2016). 


\bibitem{E3}
C. Blanco, M. Sten Delos, A. L. Erickcek, and D. Hooper, Phys. Rev. D {\bf 100}, 103010 (2019). 


\bibitem{E4}
C. Miller, A. L. Erickcek, and R. Murgia, Phys. Rev. D {\bf 100}, 123520 (2019). 


\bibitem{E5}
M. Sten Delos, T. Linden, and A. L. Erickcek, Phys. Rev. D {\bf 100}, 123546 (2019). 


\bibitem{VSEMD}
R. Allahverdi and J. K. Osiński, Phys. Rev. D {\bf 105}, 023502 (2022).


\bibitem{MAT1}
J. P. Chou, D. Curtin, and H. J. Lubatti, Phys.
Lett. B {\bf 767}, 29 (2017).


\bibitem{MAT2}
D. Curtin, M. Drewes, M. McCullough, P. Meade, R. N. Mohapatra {\it et al.}, Rept. Prog. Phys. {\bf 82}, 116201 (2019). 


\bibitem{MAT3}
C. Alpigiani {\it et al.} [MATHUSLA Collaboration], e-Print: 2009.01693 [physics.ins-det].


% \bibitem{ABCM}
% R. Allahverdi, R. Brandenberger, F-Y Cyr-Racine, and A. Mazumdar, Ann. Rev. Nucl. Part. Sci. {\bf 60}, 27 (2010). %[e-Print: arXiv:1001.2600 [hep-th]]. 


% \bibitem{Aminreview}
% M. A. Amin, M. P. Hertzberg, D. I. Kaiser, and J. Karouby, Int. J. Mod. Phys. D {\bf 24}, 1530003 (2014). %[e-Print: arXiv:1410.3808 [hep-ph]].


\bibitem{RB}
R. Allahverdi and B. Dutta, Phys. Rev. D {\bf 88}, 023525 (2013).


\bibitem{Rabi}
K. S. Babu, R. N. Mohapatra, and S. Nasri, Phys. Rev. Lett. {\bf 98}, 161301 (2007).


\bibitem{ADG}
R. Allahverdi, B. Dutta and Y. Gao, Phys. Rev. D {\bf 89}, 127305 (2014).


\bibitem{Borsanyi_gstar}
S. Borsanyi {\it et al.}, Nature {\bf 539}, 69 (2016). 


\bibitem{Kohri}
T. Hasegawa, N. Hiroshima, K. Kohri, R. S. L. Hansen, T. Tram, and S. Hannestad, JCAP {\bf 12}, 012 (2019). 


\bibitem{CMBBBN}
P. F. de Salas, M. Lattanzi, G. Mangano, G. Miele, S. Pastor,
and O. Pisanti, Phys. Rev. D {\bf 92}, 123534 (2015). 


\bibitem{ADD}
R. Allahverdi, P. S. B. Dev, and B. Dutta, Phys. Lett. B {\bf 779}, 262 (2018).


\bibitem{ADMS}
R. Allahverdi, B. Dutta, R. N. Mohapatra, and K. Sinha, Phys. Rev. Lett. {\bf 111}, 051302 (2013).


\bibitem{ADS}
R. Allahverdi, B. Dutta, and K. Sinha, Phys. Rev. D {\bf 83}, 083502 (2011).


\bibitem{Baryo}
R. Allahverdi, B. Dutta and K. Sinha, Phys. Rev. D {\bf 82}, 035004 (2010). 


\bibitem{Visible}
R. Allahverdi, B. Dutta and K. Sinha, Phys. Rev.D {\bf 87}, 075024 (2013).


\bibitem{DM}
P.S. B. Dev and R. N. Mohapatra, Phys. Rev. D {\bf 92}, 016007 (2015).


\bibitem{double1}
V. Takhistov [Super-Kamiokande Collaboration], e-Print: 1605.03235 [hep-ex].


\bibitem{double2}
 M. Litos {\it et al.}, Phys. Rev. Lett. {\bf 112}, 131803
(2014).


\bibitem{nnbar1}
M. Baldo-Ceolin {\it et al.}, Z. Phys. C {\bf 63}, 409 (1994).


\bibitem{nnbar2}
K. Abe {\it et al.} [Super-Kamiokande Collaboration], Phys.
Rev. D {\bf 91}, 072006 (2015). 
%[arXiv:1109.4227 [hep-ex]].


\bibitem{nnbar3}
B. Aharmim {\it et al.} [SNO Collaboration], e-Print: 1705.00696 [hep-ex].


\bibitem{nnbar4}
D. G. Phillips, II {\it et al.}, Phys. Rept. {\bf 612}, 1 (2016).
%[arXiv:1410.1100 [hep-ex]].


\bibitem{xenon1t}
E. Aprile {\it et al.} [XENON Collaboration], Phys. Rev. Lett. {\bf 119}, 181301 (2017).


\bibitem{pico}
C. Amole {\it et al.} [PICO Collaboration], Phys. Rev. D {\bf 100}, 022001 (2019).


\bibitem{fermi1}
M. Ackermann {\it et al.} [Fermi-LAT Collaboration], Phys. Rev. Lett. {\bf 115}, 231301 (2015).


\bibitem{fermi2}
A. Albert {\it et al.} FERMI-LAT and DES Collaborations], Astrophys. J. {\bf 834}, 110 (2017).


\bibitem{ABCO}
R. Allahverdi, I. Broeckel, M. Cicoli, and J K. Osinski, JHEP {\bf 2102}, 026 (2021).


\bibitem{FIMP}
L. J. Hall, K. Jedamzik, J. March-Russell, and S. M. West, JHEP {\bf 1003}, 080 (2010).


\bibitem{AOFI}
R. Allahverdi and J. K. Osiński, Phys. Rev. D {\bf 101}, 063503 (2020).


\bibitem{KMY}
M. Kawasaki, T. Moroi, and T. Yanagida, Phys. Lett. B {\bf 370}, 52 (1996).


\bibitem{quasidec}
I. R. Waldstein, A. L. Erickcek, and C. Ilie, Phys. Rev. D {\bf 95}, 123531 (2017).
 
 


% \bibitem{LHC1}
% B. Dutta, Y. Gao, and T. Kamon, Phys. Rev. D {\bf 89}, 096009 (2014).


% \bibitem{LHC2}
% R. Allahverdi {\it et al}, JHEP {\bf 1612}, 046 (2016).


% \bibitem{GKR}
% G. F. Giudice, E. W. Kolb, and A. Riotto, Phys. Rev. D {\bf 64}, 023508 (2001);


% \bibitem{MR}
% T. Moroi and L. Randall, Nucl. Phys. B {\bf 570}, 455 (2000).


% \bibitem{GG}
% G. B. Gelmini and P. Gondolo, Phys. Rev. D {\bf 74}, 023510 (2006).


% \bibitem{Scott1}
%  B. S. Acharya, P. Kumar, K. Bobkov, G. Kane, J. Shao,
% and S. Watson, JHEP {\bf 0806}, 064 (2008).

% \bibitem{DLS}
% B. Dutta, L. Leblond, and K. Sinha, Phys. Rev. D {\bf 80}, 035014 (2009).


% \bibitem{Scott3}
% B. S. Acharya, G. Kane, S. Watson, and P. Kumar, Phys. Rev. D 80, 083529 (2009).


% \bibitem{Beacom}
% R. K. Leane, T. R. Slatyer, J. F. Beacom, and K. C. Y. Ng, Phys. Rev. D 98, 023016 (2018).


% \bibitem{GK}
% K. Griest and M. Kamionkowski, Phys. Rev. Lett. {\bf 64}, 615 (1990).


% \bibitem{ES}
% A. L. Erickcek and K. Sigurdson, Phys. Rev. D {\bf 84}, 083503 (2011). 


% \bibitem{BR}
% G. Barenboim and J. Rasero, JHEP {\bf 1404}, 138 (2014). 


% \bibitem{FOW}
% J. J. Fan, O. Özsoy, and S. Watson, Phys. Rev. D {\bf 90}, 043536 (2014). 


% \bibitem{FPT}
% P. Fox, A. Pierce, and S. D. Thomas, e-Print: hep-th/0409059.


% \bibitem{Howie1}
% H. Baer, A. Lessa, and W. Sreethawong, JCAP {\bf 1201}, 036 (2012).


% \bibitem{Howie2}
% K. J. Bae, H. Baer, A. Lessa, and H. Serce, JCAP {\bf 1410}, 082 (2014).


% \bibitem{Dine1}
% I. Affleck and M. Dine, Nucl. Phys. B {\bf 249}, 361 (1985).


% \bibitem{Dine2}
% M. Dine, L. Randall, and S. D. Thomas, Nucl. Phys. B {\bf 458}, 291 (1996).


% \bibitem{Gravitino1}
% M. Kawasaki, K. Kohri, T. Moroi, and A. Yotsuyanagi, Phys.
% Rev. D {\bf 78}, 065011 (2008). 


% \bibitem{Gravitino2}
% R. H. Cyburt, J. Ellis, B. D. Fields, F. Luo, K. A. Olive, and V. C. Spanos, JCAP {\bf 0910}, 021 (2009).


% \bibitem{CMS}
% A. M. Sirunyan {\it et al.} [CMS Collaboration], Phys. Rev. D {\bf 97}, 092005 (2018).




\end{thebibliography}
\end{document}